\documentclass[11pt,a4paper,twoside]{article}
\usepackage[T1]{fontenc}
\usepackage[latin1]{inputenc}
\usepackage{amsmath,amssymb,amsthm}
\usepackage{graphicx,color}
\usepackage{parskip}
\def\be{\begin{equation}}
\def\ee{\end{equation}}
\def\bea{\begin{eqnarray}}
\def\eea{\end{eqnarray}}
\newcommand{\rmd}{\mbox{d}}

 \newcommand{\ket}[1]{|\kern.3ex#1\kern.3ex\rangle}
 \newcommand{\bra}[1]{\langle\kern.3ex #1 \kern.3ex|}
\begin{document}

$${\rm \bf BESSEL \;INTEGRALS,\; PERIODS\; and\; ZETA\; NUMBERS}$$

\vspace{0.5cm}

\centerline{ \bf Jean \;DESBOIS  and  St\'ephane \;OUVRY}

\vspace{0.5cm}

\centerline{LPTMS,  CNRS UMR 8626, Universit\'e Paris-Sud, 91405 Orsay Cedex}

\vspace{0.5cm}

Abstract: we present a summary of recent and older results on Bessel integrals and their relation with zeta numbers.
$${}$$
\section{INTRODUCTION}

We focus on Bessel integrals 
\be\label{one}\int_0^{\infty}{\rmd u}\;
u^{n+1}K_0(u)^{\kappa}\ee
with $n\ge 0$ and $\kappa\ge 1$ integers
and on their relation with zeta numbers.
We present   results from \cite{nous, nousbis, private, moi} and review  \cite{mash} for a self contained presentation. Continous fractions for $\zeta(3)$, $\zeta(2)$ and $\psi_1(1/3)-\psi_1(2/3)$ 
%, a number related to $\zeta(2)$, 
are presented, some of them related to (\ref{one}), others  sustained  by numerical PSLQ evidence \cite{PSLQ}. All these continuous fractions are intimately connected to Ap\'ery's continous fractions  for $\zeta(3)$ and $\zeta(2)$ irrationality demonstrations \cite{VanPoorten}. Attempts to integrate Bessel integrals starting from a multi-integral representation on a finite domain are given. In the process Bessel integrals  are shown to be periods\footnote{ Periods are defined in  \cite{Konsev} as "values of absolutely convergent integrals of rational functions with rational coefficients over domains in $R^n$ given by polynomial inequalities with rational coefficients".}. Finally  a possible way to address the irrationality of $\zeta(5)$ is proposed.

\section{REVIEW OF \cite{mash}} \label{arxiv}
%(J. Stat. Mech. (2008) P03018)}
\subsection{Quantum Mechanics and Bessel Integrals}

The random magnetic impurity model \cite{nous} describes a quantum particle in a plane coupled to a random distribution of  Aharonov-Bohm fluxes  perpendicular to the plane. It was introduced having in mind the Integer Quantum Hall effect. A perturbative expansion of the partition function of the  model in the  coupling constant $\alpha$ (the flux  expressed in  unit of the quantum of flux) for $2$ impurities, i.e. at second order  in the impurity density $\rho$,   lead us to consider Feynmann diagrams   at order $\rho^2\alpha^4$ \cite{nous} and $\rho^2\alpha^6$ \cite{private}. 

On the one hand, these Feymann diagrams, which reduce after momenta integrations to  multiple integrals on intermediate temperatures,
 were shown to rewrite in terms of simple \cite{nousbis} and double nested integrals \cite{private} on products of modified Bessel functions  
 \begin{align} I_{\rho^2\alpha^4} = & \int_0^{\infty}u\,K_0(u)^2(uK_1(u))^2 \rmd u\nonumber\\ 
I_{\rho^2\alpha^6} = & 8  \int_0^{\infty}\rmd u \,u \,K_0(u)^2(uK_1(u))^2
\int_0^{u}\rmd x\,xK_1(x)I_1(x)K_0(x)^2 \nonumber \\
&-4  \int_0^{\infty}\rmd u \, u K_0(u)(uK_1(u))\left( uK_1(u)I_0(u)-uI_1(u)K_0(u)\right)
\int_u^{\infty}\rmd x\, x\,K_0(x)^2K_1(x)^2 \nonumber\\ 
&+  \int_0^{\infty}u\,K_0(u)^4(uK_1(u))^2 \rmd u \label{wellbis}
\end{align}

On the other hand, one could show \cite {nousbis} by direct integration  
 \be\label{first0}\int_0^{\infty} u K_0(u)^4 \rmd u={2^3-1\over 8}\zeta(3)\ee 
and by integration by part that 
\be \label{first}\int_0^{\infty}u\,K_0(u)^2(uK_1(u))^2 \rmd u\ee 
is a linear combination with rational cefficients\footnote{ Linear combination  with rational coefficients   means  here that there exist three positive or negative integers $a, b$ and $c$  such that \be \nonumber a\int_0^{\infty}u\,K_0(u)^2(uK_1(u))^2 \rmd u+b\int_0^{\infty} u K_0(u)^4 \rmd u+c=0\ee } of $1$ and $\int_0^{\infty} u K_0(u)^4 \rmd u$, i.e. of $1$ and $(2^3-1)\zeta(3)$.
Similarly one obtained \cite{private,moi} by direct integration\footnote{See Appendix A for the derivations of (\ref{first0}) and (\ref{ffirst}).} 
{\be \label{ffirst}\int_{0}^{\infty}\rmd u\,  
uI_0(u)K_{0}(u)^3={2^2-1\over 8}\zeta(2)\ee} 
and by integration by part that
\be\label{firstbis} \int_0^{\infty}\rmd u\,uK_0(u)(uK_1(u))^2I_0(u)\ee
and 
\be\label{firstbis1} \int_0^{\infty}\rmd u\,uK_0(u)^2uK_1(u)uI_1(u)\ee
are linear combinations of $1$ and $\int_{0}^{\infty}\rmd u\, u 
I_0(u)K_{0}(u)^3$, i.e. of $1$ and $(2^2-1)\zeta(2)$.
Likewise,  by integration by part \cite{mash}  
\be\label{rel}\int_{0}^{\infty}u \, K_0(u)^4(uK_1(u))^2\,\rmd u =
{2\over 15}\int_{0}^{\infty}u \, K_0(u)^6\,\rmd u
-{1\over 5}\int_{0}^{\infty}u^3 \, K_0(u)^6\,\rmd u \ee 

It was then natural to argue \cite{mash} that (\ref{wellbis})  might also rewrite as a linear combination with rational coefficients of simple integrals on product of Bessel functions of weight 6 -defined as the total power of  Bessel functions- and of zeta numbers of weigth $6-1= 5$, like $\zeta(5)$, or below. The "counting rule" inferred from (\ref{first0}, \ref{first}, \ref{ffirst}, \ref{firstbis}, \ref{firstbis1}, \ref{rel}) is that an  integration $\int_0^{\infty} {\rmd u}\;u^n(\quad)$ with $n$ odd diminishes the  Bessel weight by one, so $\int_0^{\infty}u\,K_0(u)^4(uK_1(u))^2$ is like  $\zeta(5)$,  that a $I$-Bessel function  has  no weight, so $\int_{0}^{\infty}\rmd u\,  
uI_0(u)K_{0}(u)^3$  is like $\zeta(2)$ and $\int_0^{\infty}\rmd u \,u^3 \,K_0(u)^2K_1(u)^2
\int_0^{u}\rmd x\,xK_1(x)I_1(x)K_0(x)^2$  like $\zeta(3)\zeta(2)$ i.e. like $\zeta(5)$, that  in turn  is like $\int_0^{\infty}u\,K_0(u)^6$ or $\int_0^{\infty}u^3\,K_0(u)^6$ -why only odd powers  $u$ and $u^3$ appear here and no higher power will become clear later. Indeed a numerical PSLQ \cite{PSLQ} search  gave 
\be\label{PSLQ}
{I_{\rho^2\alpha^6}}=_{\rm PSLQ}{1\over 30}\int_{0}^{\infty}u \, K_0(u)^6\,\rmd u
+ {1\over 20}\int_{0}^{\infty}u^3 K_0(u)^6\,\rmd u -{2^5-1\over 160}\,\zeta(5)
\ee
(the $2^5-1$ factor multiplying $\zeta(5)$ has to be viewed in parallel with $2^3-1$ multiplying $\zeta(3)$ in (\ref{first0}) and  $2^2-1$ multiplying $\zeta(2)$ in (\ref{firstbis})). Note that
from now on   an  identity obtained from a numerical PSLQ search  will be labelled $=_{\rm PSLQ}$ as  in (\ref{PSLQ}).

The fact that the double nested integrals in (\ref{wellbis})  can be  reexpressed as linear combination with rational coefficients of simple Bessel integrals  with the right weight -here weight $6$- and of $\zeta(5)$, fits well in the  "Bessel integral  $\to$ zeta number" mapping. If
  a product  $f$ of Bessel functions   is, like in (\ref{first0}, \ref{first}, \ref{ffirst}, \ref{firstbis}, \ref{firstbis1}), mapped by simple integration on a zeta number denoted by ${\tilde \zeta}(f)$
\[f\to \int_0^{\infty}f(u)\rmd u ={\tilde\zeta}(f)\] 
%with a certain weight  
then for a pair of such  products $f,g$  the mapping by  double "nested" integration
\[f,g\to \int_0^{\infty}f(u)\rmd u \int_0^u g(x)\rmd x={\tilde \zeta}(f,g)\]
on a polyzeta number denoted by ${\tilde \zeta}(f,g)$ 
%with for weight the sum of the $\zeta(f)$ and $\zeta(g)$  weights 
makes sense, since, because of \[\int_0^{\infty}f(u)\rmd u \int_0^u g(x)\rmd x =
\int_0^{\infty}f(u)\rmd u \int_0^{\infty}
g(x) \rmd x -\int_0^{\infty}g(u)\rmd u \int_0^u f(x)\rmd x, \] 
one has
\be\label{polyfun}{\tilde\zeta}(f,g)={\tilde\zeta}(f){\tilde\zeta}(g)-{\tilde\zeta}(g,f)\ee
in analogy with
\be\label{poly}\zeta(p,q)=
\zeta(p)\zeta(q)-\zeta(p+q)-\zeta(q,p)\ee
for the standard polyzeta $\zeta(p,q)=\sum_{n>m}{1\over n^p}{1\over m^q}$ - if  $\zeta(p,q)$ would be defined as $\sum_{n>m}{1\over n^p}{1\over m^q}+{1\over 2}\zeta(p+q)$ then (\ref{poly}) would take the form  (\ref{polyfun}).

 % So far only  double nested integrals built with the $f_n$'s and  $g_m$'s have been found  to follow this pattern.

\subsection{Recurrence}

It appears that  $1$ and $ \int_{0}^{\infty}u \, K_0(u)^4\,\rmd u$ one the one hand, and  $1, \int_{0}^{\infty}u \, K_0(u)^6\,\rmd u$ and $\int_{0}^{\infty}u^3 \, K_0(u)^6\,\rmd u$ on the other hand,  play the role of  building blocks in the linear combinations reexpressing, for example, (\ref{first}) and (\ref{rel}). This might indicate their special role as basis for  more general families of Bessel integrals\footnote{The same is true of $1$ and $\int_{0}^{\infty}\rmd u\,  
uI_0(u)K_{0}(u)^3$ for  (\ref{firstbis}) and more generally for the family of Bessel integrals obtained from (\ref{injk}) by replacing either a $K_0$ by a $I_0$ or a $K_1$ by a $I_1$, see \cite{moi}.}.

Indeed,    consider such a  family of Bessel integrals  with  a given weight $\kappa$ 
\be\label{injk}
I_{n,j}^{(\kappa)} = \frac{1}{n!}\int_0^{\infty}u^{n+1}K_0(u)^{\kappa-j}K_1(u)^j\rmd u\quad \quad j=0,1,\ldots,\kappa
\ee
where $n\ge \kappa-1$ for
$I_{n,\kappa}^{(\kappa)}$ to be finite. 
Integration by parts gives
the  mapping $I_{n,\kappa}^{(\kappa)} \to I_{n+1,\kappa}^{(\kappa)}$
\be
\label{rec}I_{n,j}^{(\kappa)}={n+1\over n-j+2}
\left[ jI_{n+1,j-1}^{(\kappa)} + (\kappa-j)I_{n+1,j+1}^{(\kappa)} \right]
\ee
with the $(\kappa+1)\times (\kappa+1)$ matrix 
\be\label{matrix}
 \left(
\begin{array}{cccccccc}
0 &  \frac{\kappa(n+1)}{n+2}&0&0&0&\ldots&0&0\\
1& 0& \kappa-1&0&0&\ldots&0&0\\ 
0&  \frac{2(n+1)}{n}&0& \frac{(\kappa-2)(n+1)}{n}&0&\ldots&0&0\\ 
0&  0& \frac{3(n+1)}{n-1}& 0& \frac{(\kappa-3)(n+1)}{n-1}&\ldots& 0&0\\ 
0& 0& 0& \frac{4(n+1)}{n-2}& 0& \ldots& 0&0\\ 
\ldots& \ldots& \ldots& \ldots& \ldots& \ldots& \ldots& \ldots\\
0& 0&0&0&0& \ldots& 0&\frac{n+1}{n-\kappa+3}\\ 
0& 0&0&0&0& \ldots& \frac{\kappa(n+1)}{n-\kappa+2}&0\\ 
\end{array}\right)
\ee
 whose determinant when $\kappa$ is odd is  $-{(3)^2(5)^2\ldots(\kappa)^2(n+1)^{2\kappa+1}\over(n+2)(n+1)(n)\ldots(n-\kappa+2) }$, and when $\kappa$ is even is vanishing since in this case  
\be\label{link}
\sum_{l=0}^{\kappa/2} (-1)^l (n-2l+2)
{\kappa/2 \choose l} I_{n,2l}^{(\kappa)} = 0
\ee
Applying  (\ref{rec}) twice gives the  mapping  $n\to n+2$
\bea
I^{(\kappa)}_{n,j} & = & \frac{(n+1)(n+2)(j-1)j}{(n-j+2)(n-j+4)}\,I^{(\kappa)}_{n+2,j-2} \nonumber \\
&& + \frac{(n+1)(n+2)}{n-j+2}\left[\frac{(j+1)(\kappa-j)}{n-j+2} +
\frac{j(\kappa-j+1)}{n-j+4}\right]I^{(\kappa)}_{n+2,j} \nonumber \\
&& + \frac{(n+1)(n+2)(\kappa-j-1)(\kappa-j)}{(n-j+2)^2}\,I^{(\kappa)}_{n+2,j+2} \label{rec2}
\eea
which conserves the parity of $n-j$. It follows that  the $I^{(\kappa)}_{n,j}$'s are divided into two sub-families, depending on the parity of $n-j$.

Let us focus  on the sub-family  $n-j$  even\footnote{An analysis for the sub-family  $n-j$  odd can be done  along the same lines.}: it is enough to assume that $n$ is even, and consider 
\be\label{family1}I_{n,0}^{(\kappa)}\,,\quad I_{n,2}^{(\kappa)}\,,\quad \ldots,\quad
I_{n,\kappa}^{(\kappa)}\,,\quad n\ge \kappa\quad (\kappa \mbox{ even})\ee 
 or
\be\label{family2}I_{n,0}^{(\kappa)}\,,\quad I_{n,2}^{(\kappa)}\,,\quad \ldots,\quad
I_{n,\kappa-1}^{(\kappa)}\,,\quad n\ge \kappa-1\quad (\kappa \mbox{ odd})\ee
and  note that 
\be\label{fa1} I_{n+1,1}^{(\kappa)}\,,\quad I_{n+1,3}^{(\kappa)}\,,\quad \ldots, \quad I_{n+1,\kappa-1}^{(\kappa)}
\quad (\kappa \mbox{ even})\ee
or
\be\label{fa2} I_{n+1,1}^{(\kappa)}\,,\quad I_{n+1,3}^{(\kappa)}\,, \quad \ldots, \quad I_{n+1,\kappa}^{(\kappa)}
\quad (\kappa \mbox{ odd})\ee
are respectively related to (\ref{family1}) and (\ref{family2})
by inverting (\ref{rec}), keeping in mind  (\ref{link}) when $\kappa$  is even\footnote{
The integrals  (\ref{family1}) and (\ref{fa1})  (respectively (\ref{family2}) and (\ref{fa2})) are
tantamount to the set
\be\label{set}\int_0^{\infty}u^{n+1} K_0(u)^{\kappa-j} (uK_1(u))^{j} \, \rmd u\quad\quad n \;{\rm even}\;\ge 0\ee with $\kappa$ even (respectively $\kappa$ odd).}.

By inverting  (\ref{rec2}) (again  keeping in mind  (\ref{link})), all integrals in (\ref{family1}) (and thus in (\ref{fa1})) 
are  linear combinations with rational coefficients of the initial conditions
 \be\label{ouf1}\{I_{\kappa,0}^{(\kappa)},\quad I_{\kappa,2}^{(\kappa)},\quad \ldots,\quad I_{\kappa,\kappa}^{(\kappa)}\}\quad (\kappa \mbox{ even})\ee 
Now,  on the one hand  from (\ref{rec})  one has 
$2I_{\kappa-1,\kappa-1}^{(\kappa)}={\kappa}(I_{\kappa,\kappa}^{(\kappa)}+(\kappa-1) I_{\kappa,\kappa-2}^{(\kappa)})$
and, on the other hand,  $I_{\kappa-1,\kappa-1}^{(\kappa)}=1/\kappa!$. It follows that 
  $I_{\kappa,\kappa}^{(\kappa)}$ can be replaced by $1$.
Using (\ref{link}), one can drop one more  element:
therefore, all integrals in (\ref{family1}) (and in (\ref{fa1})) are linear combinations
with rational coefficients  of the $\kappa/2$  numbers
 
\be\label{bkeven}
\{1\,,\quad I_{\kappa,0}^{(\kappa)}\,,\quad  I_{\kappa,2}^{(\kappa)}\,,\quad \ldots, \quad I_{\kappa,\kappa-4}^{(\kappa)}\}\quad (\kappa \mbox{ even})
\ee

Likewise, all integrals in (\ref{family2}) (and thus in (\ref{fa2})) 
are  linear combinations with rational coefficients of
\be\label{ouf2}\{I_{\kappa-1,0}^{(\kappa)},\quad I_{\kappa-1,2}^{(\kappa)}, \quad\ldots, \quad I_{\kappa-1,\kappa-1}^{(\kappa)}\}\quad (\kappa \mbox{ odd})\ee
that is to say, since $I_{\kappa-1,\kappa-1}^{(\kappa)}=1/\kappa!$, of
the $(\kappa+1)/2$  numbers
\be\label{bkodd}
\{1\,,\quad I_{\kappa-1,0}^{(\kappa)}\,, \quad I_{\kappa-1,2}^{(\kappa)}\,,\quad \ldots,\quad I_{\kappa-1,\kappa-3}^{(\kappa)}\}
\quad (\kappa \mbox{ odd})\ee

Finally, by applying (\ref{rec}) appropriately
for $0\le n\le \kappa$,    (\ref{bkeven}) can be mapped on  
\be\label{ba1} \{1\,,\quad I_{0,0}^{(\kappa)}\,,\quad I_{2,0}^{(\kappa)}\,,\quad
 \ldots, \quad I_{\kappa-4,0}^{(\kappa)}\}\quad (\kappa \mbox{ even})\ee
and   (\ref{bkodd})
 on
\be\label{ba2}\{ 1\,,\quad I_{0,0}^{(\kappa)}\,, \quad I_{2,0}^{(\kappa)}\,,\quad
 \ldots,\quad I_{\kappa-3,0}^{(\kappa)}\}\quad (\kappa \mbox{ odd})\ee
 
It follows that for $\kappa$ even the $\kappa/2$ numbers in (\ref{ba1}), namely  
%\be\label{ba1bis} \{1\,,\quad\int_0^{\infty}u^{}K_0(u)^{\kappa}\rmd u \,,\quad {1\over 2!}\int_0^{\infty}u^{3}K_0(u)^{\kappa}\,,\quad
% \ldots,\quad{1\over (\kappa-4)!}\int_0^{\infty}u^{\kappa-3}K_0(u)^{\kappa} \}\ee
\be\label{ba1bis} \{1\,,\quad\int_0^{\infty}u^{}K_0(u)^{\kappa}\rmd u \,,\quad \int_0^{\infty}u^{3}K_0(u)^{\kappa}\,,\quad
 \ldots,\quad{}\int_0^{\infty}u^{\kappa-3}K_0(u)^{\kappa} \}\ee
constitute  a basis for the integrals (\ref{family1}) and (\ref{fa1}). 
Likewise for $\kappa$ odd, the $(\kappa+1)/2$ numbers  in  (\ref{ba2}), namely 
%\be\label{ba2bis} \{1\,,\int_0^{\infty}u^{}K_0(u)^{\kappa}\rmd u \,, {1\over 2!}\int_0^{\infty}u^{3}K_0(u)^{\kappa}\,,
 %\ldots,{1\over (\kappa-3)!}\int_0^{\infty}u^{\kappa-2}K_0(u)^{\kappa} \}\ee
 \be\label{ba2bis} \{1\,,\int_0^{\infty}u^{}K_0(u)^{\kappa}\rmd u \,, \int_0^{\infty}u^{3}K_0(u)^{\kappa}\,,
 \ldots,\int_0^{\infty}u^{\kappa-2}K_0(u)^{\kappa} \}\ee
constitute  a basis for the integrals (\ref{family2}) and (\ref{fa2}). 

By basis one means that these numbers should be independent over $Q$: none of them is a linear combination with rational coefficients
of the others i.e. there are no positive or negative integers $a, a_n$ such that
\be a\times 1+\sum_{n\;{\rm even}=0}^{\kappa-4} a_n \times \int_0^{\infty}u^{n+1}K_0(u)^{\kappa}=0\quad (\kappa \mbox{ even})\ee  or
  such that
\be a\times 1+\sum_{n\;{\rm even}=0}^{\kappa-3} a_n\times \int_0^{\infty}u^{n+1}K_0(u)^{\kappa}=0\quad (\kappa \mbox{ odd})\ee 
This implies  that the numbers in (\ref{ba1bis}) with $\kappa$ even  
 are irrational  and also irrational the one relatively to the other. The same should be true of the numbers in (\ref{ba2bis}) with  $\kappa$  odd.
\subsection{Asymptotic eigenvalues}
 In   the asymptotics limit  $n\to\infty$,  the mapping (\ref{rec2}) becomes 
\be
I^{(\kappa)}_{n,j} = (j-1)jI^{(\kappa)}_{n+2,j-2} + (2j(\kappa-j)+\kappa)I^{(\kappa)}_{n+2,j}
+(\kappa-j-1)(\kappa-j)I^{(\kappa)}_{n+2,j+2} 
\ee
The $[(\kappa - \kappa\bmod2)/2 + 1]$ eigenvalues of the resulting matrix  are
$\kappa^2$, $(\kappa-2)^2$, $(\kappa-4)^2$, \ldots\, where
the last eigenvalue is 1 for odd $\kappa$ and 0 for even $\kappa$.
In the latter case one has to reduce the dimension of the matrix by one unit, using (\ref{link}).
Thereupon, the largest eigenvalue of the inverse matrix is 1 or 1/4 for
$\kappa$ odd or even, respectively, whereas the smallest one is $1/\kappa^2$.
Remarkably, with the
initial conditions (\ref{ouf1}),~(\ref{ouf2}), it is the
smallest eigenvalue $1/\kappa^2$ that determines the asymptotic behavior of the   inverse matrix\footnote{Starting from any other initial condition would lead to an asymptotics governed by the highest eigenvalue.}.
This can be understood by noting that
the eigenvector corresponding to the  eigenvalue $\kappa^2$ is $\{1,1,\ldots,1\}$,  since $ (j-1)\times 1+ (2j(\kappa-j)+\kappa)\times 1
+(\kappa-j-1)(\kappa-j)\times 1= \kappa^2$ for all $j$'s.
Now, the $I^{(\kappa)}_{n\to\infty,j}$'s happen to
not depend on $j$, because their integrand $u^{n+1}K_0(u)^{\kappa-j}K_1(u)^j$
peaks when $n\to\infty$ at large values of $u$, where both $K_0(u)$ and $K_1(u)$
are approximated by
$K_{\nu}(u)
\raisebox{-0.53em}{$\stackrel{\textstyle\longrightarrow}{\scriptstyle u\to\infty}$}
\sqrt{{\pi\over 2u}}e^{-u}$.
Therefore,  in the asymptotic limit, the vector $\{I_{n\to\infty,j}^{(\kappa)}\}$ is indeed
proportional to $\{1,1,\ldots,1\}$.
%This consideration is of course by itself not sufficient: it is only when
%the  initial conditions match a certain  unique condition
%that the smallest eigenvalue indeed dominates.

\subsection{Question}\label{question}
 
The irrationality claims on (\ref{ba1bis}) and (\ref{ba2bis})  have  been  checked numerically.  They are also supported by the  eigenvalues   discussion above:  (\ref{ouf1})  and (\ref{ouf2}) uniquely
lead to an  asymptotic behavior governed by the corresponding  smallest  eigenvalue  and are decomposed
on the basis (\ref{ba1bis}) and (\ref{ba2bis}). It might
mean that the building blocks of (\ref{ba1bis}) and (\ref{ba2bis}),
 Bessel integrals $\int_0^{\infty}u^{n+1}\, K_0(u)^{\kappa}\,\rmd u$
with $n$ even, play some special role in number theory.
Clearly the cases  $\kappa=1$ and $\kappa=2$ are trivial since  $\int_0^{\infty}u^{n+1}\, K_0(u)^{1-j}(uK_1(u))^j\,\rmd u$ with $j=0,1$  and $\int_0^{\infty}u^{n+1}\, K_0(u)^{2-j}(uK_1(u))^{j}\,\rmd u$ with $j=0,1,2$ are all rational  and accordingly  both the basis (\ref{ba2}) for $\kappa=1$ and (\ref{ba1})  for $\kappa=2$  have for sole element  $1$.
So  the question asked when $\kappa\ge 3$: is it possible     to assess the irrationality of the basis (\ref{ba1bis}) and (\ref{ba2bis})?

One will show in the next section that in the first non trivial  cases $\kappa=4$ and $\kappa=3$  where the basis  (\ref{ba1}) and (\ref{ba2}) have dimension  $2$ the algebra above  leads to continous fractions  with seemingly insufficient fast numerical convergence to hint at the irrationality of $I_{0,0}^{(4)}$  or $I_{0,0}^{(3)}$.

As far as the case $\kappa=4$ is concerned,  this question is formal  since $I_{0,0}^{(4)}$  can be integrated to a rational number times $\zeta(3)$, which is known to be irrational \cite{VanPoorten}. When $\kappa=3$ on the other hand,  $I_{0,0}^{(3)}$ can be also be integrated  to  a rational number times  $\psi_1(1/3)-\psi_1(2/3)$, a $\zeta(2)$-like number, whose irrationality is so far not known.  Some intimate relation with Ap\'ery's proofs of the irrationality of $\zeta(3)$ and $\zeta(2)$  will show up in the process. 

\section{CONTINUOUS FRACTIONS}\label{goodsection}

\subsection{ Weight $\kappa=4$  $\to \zeta(3)$}

The $n\to n+2$ mapping (\ref{rec2})
\[\left(
\begin{array}{c}
I_{n,0}^{(4)}\\ 
I_{n,2}^{(4)}\\  I_{n,4}^{(4)}\end{array}\right)
={2(1 + n)}
\left(
\begin{array}{ccc}
{2 \over 2 + n} & {6 \over 2 + n}&0\\  
  {1\over n}& {6(1 + n)\over n^2}&{ (2+n)\over n^2}\\
 0 & {6(2+n)\over (-2 + n) n}&{ 2 (2+n)\over (-2 + n) n}  
\end{array}\right)
\left(
\begin{array}{c}
I_{n+2,0}^{(4)}\\I_{n+2,2}^{(4)}\\I_{n+2,4}^{(4)} 
\end{array}\right)
%\nonumber\ee
\]
leads to, using  $(n + 2)I_{n,0}^{(4)}- 2 n I_{n,2}^{(4)}+ (n - 2)I_{n,4}^{(4)}=0$ in (\ref{link})
\be\label{simple0}\left(
\begin{array}{c}
I_{n,0}^{(4)}\\ 
I_{n,4}^{(4)}\end{array}\right)
=(1 + n)
\left(
\begin{array}{cc}
{ 32 + 10n \over (2 + n)^2} & {6 n \over(2 + n)^2}\\  
{   6 (4 + n) \over(-2 + n) n}&{ 8 + 10 n\over(-2 + 
     n) n}  
\end{array}\right)
\left(
\begin{array}{c}
I_{n+2,0}^{(4)}\\ 
I_{n+2,4}^{(4)}\end{array}\right)
%\nonumber\ee
\ee
with asymptotics when $n\to\infty$
\[
\left(
\begin{array}{cc}
{ 10 }& {6 }\\  
   {6 }&{  10  } 
\end{array}\right)=\left(
\begin{array}{cc}
 {\lambda_++\lambda_-\over 2 }& {\lambda_+-\lambda_-\over 2  }\\  
   {\lambda_+-\lambda_-\over 2  }&  {\lambda_++\lambda_-\over 2}   
\end{array}\right)=\left(
\begin{array}{cc}
{ 1 }& {1  }\\  
   {1  }&{  -1}    
\end{array}\right)\left(
\begin{array}{cc}
{ \lambda_+ }& {0  }\\  
   {0  }&{  \lambda_-  } 
\end{array}\right)\left(
\begin{array}{cc}
{ 1} & {1  }\\  
   {1  }&{  -1}    
\end{array}\right)^{-1}\]
and eigenvalues $\{\lambda_-=4,\lambda_+=16\}$ and eigenvectors $\{\{1,-1\}, \{1,1\}\}$.
As already stated in the Introduction the asymptotic scaling is governed by the asymptotic   eigenvector $ \{1,1\}$ corresponding to the eigenvalue $\lambda_+=16$ (and thus to $1\over 16$  for the inverse iteration).
%\[\left(
%\begin{array}{c}
%I_{2k,0}^{(4)}\\ 
%I_{2k,4}^{(4)}\end{array}\right)
%=(1 + 2 k)
%\left(
%\begin{array}{cc}
%{ (8 + 5 k)\over (1 + k)^2} & {3 k \over(1 + k)^2}\\  
%   {3 (2 + k) \over(-1 + k) k}&{ (2 + 5 k)\over(-1 + 
%     k) k} } 
%\end{array}\right)
%\left(
%\begin{array}{c}
%I_{2(k+1),0}^{(4)}\\ 
%I_{2(k+1),4}^{(4)}\end{array}\right)
%\nonumber\ee
%\]
Since $n$ is even set $n=2k$ and define
\be u(k)={I_{2k,4}^{(4)}\over I_{2k,0}^{(4)}}\ee
to obtain
%$$u(k)={3(k+2)(k+1)^2+(5k+2)(k+1)^2u(k+1)\over 
% k(k-1)(5k+8) +3k^2(k-1)u(k+1) }$$
%which is  of the form 
$$u(k)={a(k)+c(k)u(k+1)\over 
 b(k) +d(k)u(k+1) }={c(k)\over 
 d(k) } +{a(k)d(k)-b(k)c(k)\over d(k)(b(k) +d(k)u(k+1))} $$
with 
\begin{align}a(k)&=3 (2 + k) (1+k)^2\nonumber\\
 b(k)&=(8 + 5 k)(-1 + k) k\nonumber\\
 c(k)&=(2 + 5 k)(1+k)^2\nonumber\\
 d(k)&=3k(-1 + k) k\label{abcd}\end{align}
%$$d(k)u(k)-c(k)= 
 %{a(k)d(k)-b(k)c(k)\over d(k)}{1\over {b(k)\over d(k)} +{d(k+1)u(k+1)-c(k+1)+c(k+1)\over d(k+1)} $$
 Call
 $$d(k)u(k)-c(k)=z(k)$$
 and get the  iteration $[z(k)\to z(k-1)]$ with the  continuous fraction
 $$z(k)={(a(k)d(k)-b(k)c(k))d(k+1)\over d(k)}{1\over d(k+1)({b(k)\over d(k)}+{c(k+1)\over d(k+1)}) +z(k+1)} $$
This is finally 
 \be\label{fraction}z(k-1)={-16 k^6\over 2+9k+15k^2+10k^3+z(k)}\ee
 with $k\to\infty$ asymptotics 
\[z(k)\simeq{-16 k^6\over 10k^3+z(k)}\Rightarrow \lim_{k\to\infty}{z(k)\over k^3}=\{-2, -8\}\] 
here $\lim_{k\to\infty}{z(k)/k^3}=\lim_{k\to\infty} {(d(k)u(k)-c(k))/ k^3}=-2$ since one iterates from   $u(\infty)=1$. 
 Calling ${\tilde{z}}(k)=d(k)({b(k-1)\over d(k-1)}+{c(k)\over d(k)}) +z(k)=d(k)(u(k)+{b(k-1)\over d(k-1)}))$, the inverse iteration
 % $[{\tilde{z}}(k-1)\to {\tilde{z}}(k)]$
  is
 \[{\tilde{z}}(k)={-16 k^6\over 2-9k+15k^2-10k^3+{\tilde{z}}(k-1)}\]
  with %\footnote{Or, in view of  a better grasp on the asymptotic behavior altogether with  a finite $k\to\infty$ scaling, call rather  $$z'(k)={z(k)\over d(k)}=u(k)-{c(k)\over d(k)}\Rightarrow z'(k)={(a(k)d(k)-b(k)c(k))\over d(k)^2}{1\over {b(k)\over d(k)}+{c(k+1)\over d(k+1)} +z'(k+1)}$$  with asymptotics $z'(\infty)=u(\infty)-{c(\infty)\over d(\infty)}=\pm 1 -{{\lambda_+-\lambda_-\over 2}\over{\lambda_++\lambda_-\over 2}}=\{-2{\lambda_-\over \lambda_++\lambda_-},-2{\lambda_+\over \lambda_++\lambda_-}\}=\{-{2\over 3},-{8\over 3}\}$ (or directly ${{z}}'(\infty)=-{16 \over 9 }{1\over {10\over 3} +{{z}}'(\infty)}\Rightarrow {{z}}'(\infty)=-{1\over 3}\{{2}, {8}\}$). As already known one iterates from  $u(\infty)=1$ so the asymptotics  is  $-{2\over 3}$. Call now ${\tilde{z}}'(k)=z'(k)+{b(k-1)\over d(k-1)}+{c(k)\over d(k)}=u(k)+{b(k-1)\over d(k-1)}$: the inverse iteration is then $${\tilde{z}}'(k+1)={(a(k)d(k)-b(k)c(k))\over d(k)^2}{1\over -({b(k-1)\over d(k-1)}+{c(k)\over d(k)}) +{\tilde{z}}'(k)} $$ i.e. $${\tilde{z}}'(k)={-16 k^4\over 9 ( k-2) ( k-1)^3 }{1\over -{(2 k - 1) (5 k^2 - 5 k + 2)\over 3 (k - 2) (k - 1)^2} +{\tilde{z}}'(k-1)} $$ with initial condition ${\tilde{z}}'(2)=u(2)+{b(1)\over d(1)}=u(2)+{13\over 3}$ and asymptotics ${\tilde{z}}'(\infty)=u(\infty)+{b(\infty)\over d(\infty)}=\pm 1+{{\lambda_+-\lambda_-\over 2}\over{\lambda_++\lambda_-\over 2}}=\{2{\lambda_+\over \lambda_++\lambda_-},2{\lambda_-\over \lambda_++\lambda_-}\}=\{{8\over 3},{2\over 3}\}$ (or directly  ${\tilde{z}}'(\infty)=-{16 \over 9 }{1\over -{10\over 3} +{\tilde{z}}'(\infty)}\Rightarrow {\tilde{z}}'(\infty)={1\over 3}\{2, 8\}$). As already known here  $u(2)= (\ref{original})$   leads  to the   asymptotics ${8\over 3}$ with $u(\infty)=1$ and any  $u(2)\ne(\ref{original})$ would lead to   ${2\over 3}$ with $u(\infty)=-1$.} 
asymptotics 
\[{\tilde{z}}(k)\simeq {-16 k^6\over -10k^3+{\tilde{z}}(k)}\Rightarrow \lim_{k\to\infty}{{\tilde{z}}(k)\over k^3}=\{2, 8\}\] 
here $\lim_{k\to\infty}{{\tilde{z}}(k)/k^3}=\lim_{k\to\infty}{(d(k)u(k)+b(k))/ k^3}= 8$ since one iterates from  the  Bessel integrals initial condition ${\tilde{z}}(2)$. 

Continuous fractions appear in  Ap\'ery's  proof of  $\zeta(3)$  (and  $\zeta(2)$) irrationality:
 
 \be\label{apery}\zeta(3)={6+0()\over 5+1\bigg({-1^6\over P(1)+{-2^6\over P(2)+{-3^6\over P(3)+...}}}\bigg)}\ee
 with the Ap\'ery polynomial
 $$P(k)=5+27k+51k^2+34k^3=(2k+1)(17k^2+17k+5)$$
 This corresponds to the iteration
 
 \be z(k-1)={- k^6\over P(k)+z(k)}\label{coucou}\ee
 
and to the rational approximation 
 \be\label{fast}\zeta(3)=\lim_{k\to\infty}{{y(k)}_{\{y(0)=0,y(1)=6\}}\over y(k)_{\{y(0)=1,y(1)=5\}}}\ee
  
with

\be\label{fractionun}y(k+1)-P(k)y(k)+k^6y(k-1)=0\ee

The initial conditions  are i.e.
 $y(0)=0, \;y(1)=6\to  y(k)_{\{y(0)=0,y(1)=6\}}$  
 and
 $y(0)=1,\;y(1)=5\to  y(k)_{\{y(0)=1,y(1)=5\}}$.
  $\zeta(3)$ is proven to be irrational
thanks to the  sufficiently fast convergence with respect to the increasing size of the denominators of the rational approximation (\ref{fast}).  
  
 In the weight $\kappa=4$ case the iteration (\ref{fraction}) starts at $k=2$ ($n=4$). From (\ref{fraction}) 

\be z(2)={-16\times 3^6\over 2+9k+15k^2+10k^3|_{k=3} +{-16 \times 4^6\over 2+9k+15k^2+10k^3|_{k=4}+...}}=\lim_{k\to\infty}{{y(k)}_{\{y(2)=1,y(3)=0\}}\over y(k)_{\{y(2)= 0,y(3)= 1\}}}\label{approx}\ee
with
\be\label{fractionbis} y(k+1)-({2+9k+15k^2+10k^3})y(k)+16k^6y(k-1)=0\ee
Using (\ref{rec}) the $k=2$ ($n=4$) initial conditions  are reexpressed  on the basis (\ref{ba1bis}) for $\kappa=4$, i.e. on  $\{1,\quad I_{0,0}^{(4)}\}$, as
 \[I_{4,0}^{(4)}=-{9\over 512}\times 1+{7\over 384}\times{I_{0,0}^{(4)}}\quad{\rm and} \quad
I_{4,4}^{(4)}={53\over1536}\times 1-{3\over128}\times {I_{0,0}^{(4)}}\] 
so
% \be\label{original}  u(2)={53-36{I_{0,0}^{(4)}}\over -27+28{I_{0,0}^{(4)}}}\Rightarrow z(2)=-96{37-36I_{0,0}^{(4)}\over 27 -28 %I_{0,0}^{(4)}}\ee  
% and therefore
\be\label{originalbis} z(2)=-96{37 -36 I_{0,0}^{(4)}\over 27-28 I_{0,0}^{(4)}}=\lim_{k\to\infty}{{y(k)}_{\{y(2)=1,y(3)=0\}}\over y(k)_{\{y(2)= 0,y(3)= 1\}}}\ee

It appears numerically that the convergence of the  rational approximation (\ref{approx})  does not seem  sufficiently fast to prove the irrationality of $z(2)$ that is to say that of $I_{0,0}^{(4)}$.
%\end{itemize}
 One notes in (\ref{fraction}) 
\begin{itemize}
\item 
 in the denominator  
 $2+9k+15k^2+10k^3=P(k)-3(2k+1)^3$
 where
  $P(k)$ 
 is the Ap\'ery polynomial.
\item in
  the numerator   $k^6$  as in the Ap\'ery case.
  \end{itemize}
  
These are not coincidences:  $I_{0,0}^{(4)}=\int_0^{\infty}uK_0(u)^4\,\rmd u$ has already been integrated in  (\ref{first0}) to be proportionnal to $\zeta(3)$ 
   \be\label{sign} I_{0,0}^{(4)}={2^3-1\over 8}\zeta(3)={\sum_{p=1}^{\infty}{1\over p^3}-\sum_{p=1}^{\infty}{(-1)^p\over p^3}\over 2}\ee
so that 
%(\ref{fraction}, \ref{fractionbis}) and 
 (\ref{originalbis}) is in fact
a rational approximation to $\zeta(3)$  
%\[z(k-1)={-16 k^6\over P(k)-3(2k+1)^3+z(k)}\]

%$$y(k+1)-({P(k)-3(2k+1)^3})y(k)+16k^6y(k-1)=0$$

$$ z(2)=-96{74-63{{\zeta}}(3)\over 54-49{{\zeta}}(3)}=\lim_{k\to\infty}{{y(k)}_{\{y(2)=1,y(3)=0\}}\over y(k)_{\{y(2)= 0,y(3)= 1\}}}$$

One can push  further the analogy by rewriting both Ap\'ery   and weight $\kappa=4$ rational approximations to $\zeta(3)$   starting  from  $z(0)$ i.e.   with  initial  conditions 
 $\{y(0)=1, \;y(1)=0\}$
 and  $\{y(0)=0, \; y(1)=1\}$.
 
 {\bf  Ap\'ery:} from
 
% $$z(k-1)={- k^6\over P(k)+z(k)}$$
 
 $$\zeta(3)={6+0()\over 5+1\bigg({-1^6\over P(1)+{-2^6\over P(2)+{-3^6\over P(3)+...}}}\bigg)}=\lim_{k\to\infty}{{y(k)}_{\{y(0)=0,y(1)=6\}}\over y(k)_{\{y(0)=1,y(1)=5\}}}$$
 
one deduces 
 $$ z(0)={6\over \zeta(3)}-5={-1^6\over P(1)+{-2^6\over P(2)+{-3^6\over P(3)+...}}}=\lim_{k\to\infty}{y(k)_{\{y(0)=1,y(1)=0\}}\over y(k)_{\{y(0)=0,y(1)=1\}}}$$

with $y(k)$ given in (\ref{fractionun}).

 {\bf Weight $\kappa=4$:}  one starts from $z(2)$ in (\ref{originalbis}) and with (\ref{fraction}) iterates to

%$$ z(2)=-96{37-36 I_{0,0}^{(4)}\over 27-28 I_{0,0}^{(4)}}\Rightarrow z(1) = -{4 (-27 + 28I_{0,0}^{(4)})\over -3 + 4I_{0,0}^{(4)}}$$

\be\label{simple} z(0) =
%-{2 (-3 + 4I_{0,0}^{(4)})\over 4I_{0,0}^{(4)}}=
{3\over 2I_{0,0}^{(4)}}-2
%={12 \over 7\zeta(3)}-2
\ee
where  a simplification has occured since  one would have expected in general
\[z(0)=
 {a +b\;I_{0,0}^{(4)}\over c+d\;I_{0,0}^{(4)}}\]

Finally one obtains
%$$z(0)={12 \over 7\zeta(3)}-2
%={-16\times 1^6\over P(1)-3(2\times 1+1)^3+{-16 \times 2^6\over P(2)-3(2\times 2+1)^3+{-16 \times 3^6\over P(3)-3(2\times %3+1)^3...}}}
%=\lim_{k\to\infty}{{y(k)}_{\{y(0)= 1,y(1)= 0\}}\over y(k)_{\{y(0)= 0,y(1)= 1\}}}$$
$$z(0)={3\over 2I_{0,0}^{(4)}}-2={12 \over 7\zeta(3)}-2
=\lim_{k\to\infty}{{y(k)}_{\{y(0)= 1,y(1)= 0\}}\over y(k)_{\{y(0)= 0,y(1)= 1\}}}$$
with $y(k)$ given in (\ref{fractionbis}).

So  finally one has 

\begin{itemize}

\item {\bf  Ap\'ery:}

$$z(k-1)={- k^6\over P(k)+z(k)}={- k^6\over (2k+1)(17k^2+17k+5)+z(k)}$$

with asymptotics 

$$z(k)\simeq {- k^6\over 34k^3+z(k)}\Rightarrow\lim_{k\to\infty} {z(k)\over k^3}=\{-(1+\sqrt{2})^4,\;-{\displaystyle 1\over\displaystyle (1+\sqrt{2})^{4}}\}$$

$$y(k+1)-P(k)y(k)+k^6y(k-1)=0\to_{k\to\infty} y^2-34 y+1=0\to \{(1+\sqrt{2})^4,\; {\displaystyle 1\over\displaystyle (1+\sqrt{2})^{4}}\}$$

$$z(0)={42\over 7\zeta(3)}-5=\lim_{k\to\infty}{{y(k)}_{\{y(0)= 1,y(1)= 0\}}\over y(k)_{\{y(0)= 0,y(1)= 1\}}}$$

\item{\bf  Weight\footnote{One could use  that $2+9k+15k^2+10k^3=P(k)-3(2k+1)^3=(1+2k)(5k^2+5k+2)$  is a multiple of 2 : then
\be\label{notrebis}z(k-1)={-4 k^6\over {P(k)-3(2k+1)^3\over 2}+z(k)}={-4 k^6\over (2k+1){5k^2+5k+2\over 2}+z(k)}\ee
$$y(k+1)-{P(k)-3(2k+1)^3\over 2}y(k)+4k^6y(k-1)=0\to y^2-5 y+4=0\to \{4,\;1\}$$
$$z(0)={6 \over 7\zeta(3)}-1=\lim_{k\to\infty}{{y(k)}_{\{y(0)= 1,y(1)= 0\}}\over y(k)_{\{y(0)= 0,y(1)= 1\}}}$$} $\kappa=4$:}

\be\label{notreter}z(k-1)={-16 k^6\over {P(k)-3(2k+1)^3}+z(k)}={-16 k^6\over (2k+1)(5k^2+5k+2)+z(k)}\ee

with asymptotics 

$$z(k)\simeq{-16 k^6\over 10k^3+z(k)}\Rightarrow \lim_{k\to\infty}{z(k)\over k^3}=\{-8,\; -2\}$$

$$y(k+1)-({P(k)-3(2k+1)^3})y(k)+16k^6y(k-1)=0\to_{k\to\infty} y^2-10 y+16=0\to \{8,\; 2\}$$

$$z(0)={12 \over 7\zeta(3)}-2=\lim_{k\to\infty}{{y(k)}_{\{y(0)= 1,y(1)= 0\}}\over y(k)_{\{y(0)= 0,y(1)= 1\}}}$$

\end{itemize}

 A PSLQ search  also gave

$$z(k-1)={- k^6\over {P(k)-2(2k+1)^3\over 3}+z(k)}={- k^6\over (2k+1)(3k^2+3k+1)+z(k)}$$

with asymptotics 

$$z(k)\simeq{- k^6\over 6k^3+z(k)}\Rightarrow\lim_{k\to\infty}{z(k)\over k^3}=\{-(1+\sqrt{2})^2,\; -{\displaystyle 1\over\displaystyle (1+\sqrt{2})^{2}}\}$$

$$y(k+1)-{P(k)-2(2k+1)^3\over 3}y(k)+k^6y(k-1)=0\to_{k\to\infty} y^2-6 y+1=0\to \{(1+\sqrt{2})^2,\;{\displaystyle 1\over\displaystyle (1+\sqrt{2})^{2}}\}$$

$$z(0)=_{\rm PSLQ}{ 8\over 7\zeta(3)}-1=\lim_{k\to\infty}{{y(k)}_{\{y(0)= 1,y(1)= 0\}}\over y(k)_{\{y(0)= 0,y(1)= 1\}}}$$

(one has used  that $P(k)-2(2k+1)^3=(2k+1)(9k^2+9k+3)$ is a multiple of 3).

Only the Ap\'ery rational approximation has a sufficiently fast convergence  to check numerically  the irrationality of $z(0)={6\over\zeta(3)}-5$, i.e. of ${\zeta(3)}$.
 
\subsection{Weight $\kappa=3\to \psi_1(1/3)-\psi_1(2/3)$}

The $n\to n+2$ mapping (\ref{rec2}) is \[\left(
\begin{array}{c}
I_{n,0}^{(3)}\\ 
I_{n,2}^{(3)}\end{array}\right)
={(1+n)}
\left(
\begin{array}{cc}
{3 \over 2 + n} & {6 \over 2 + n}\\  
 {2\over n}& {(6+7n)\over n^2}
\end{array}\right)
\left(
\begin{array}{c}
I_{n+2,0}^{(3)}\\I_{n+2,2}^{(3)} 
\end{array}\right)\]
with asymptotics matrix $n\to\infty$
\[
\left(
\begin{array}{cc}
{ 3 }& {6 }\\  
   {2 }&{  7  } 
\end{array}\right)=\left(
\begin{array}{cc}
{ \lambda_++3\lambda_-\over 4} & {3(\lambda_+-\lambda_-)\over 4 }\\  
   {\lambda_+-\lambda_-\over 4}&{  3\lambda_++\lambda_-\over 4}   
\end{array}\right)=\left(
\begin{array}{cc}
{ 1 }& {1  }\\  
   {1  }&-{1\over 3   } 
\end{array}\right)\left(
\begin{array}{cc}
{ \lambda_+} & {0  }\\  
   {0  }&  {\lambda_-}   
\end{array}\right)\left(
\begin{array}{cc}
{ 1} & {1  }\\  
   {1  }& -{1\over 3}  
\end{array}\right)^{-1}\]
with eigenvalues $\{\lambda_-=1,\lambda_+=9\}$ and eigenvectors $\{\{1,-1/3\}, \{1,1\}\}$. As already stated in the Introduction the asymptotic scaling is governed by the asymptotic   eigenvector $ \{1,1\}$ corresponding to the eigenvalue $\lambda_+=9$ (and thus to $1\over 9$  for the inverse iteration).

 %\[\left(
%\begin{array}{c}
%q_{k}(0)\\ 
%q_{k}(1)\end{array}\right)
%={(1 + 2 k)}
%\left(
%\begin{array}{cc}
%{3\over 2+2k} & {3\over 1+k}\\  
% {1\over k} & {3+7k\over 2k^2} 
%\end{array}\right)
%\left(
%\begin{array}{c}
%q_{k+1}(0)\\ 
%q_{k+1}(1)\end{array}\right)
%\nonumber\ee
%\]
Define 
$$u(k)={I_{2k,2}^{(3)}\over I_{2k,0}^{(3)}}$$ 
call
 \[z(k)=d(k)u(k)-{c(k)}\]
 and obtain, using
 \[a(k)=2k(1+k)\]
 \[b(k)=3k^2\]
 \[c(k)=(1+k)(3+7k)\]
 \[d(k)=6k^2\]
 the iteration $[z(k)\to z(k-1)]$ with  the continuous fraction
 \be\label{rec3} z(k-1)={-9k^4\over 10k^2+10k+3+z(k)}\ee
with asymptotics  
 \[z(k)\simeq {-9k^4\over 10k^2+z(k)}\Rightarrow \lim_{k\to\infty}{z(k)\over k^2}=\{-1, -9\}\] 
here $-k^2$ since one iterates from   $u(\infty)=1$.
 Calling ${\tilde{z}}(k)= 2+9k+15k^2+10k^3+z(k)$, the inverse   iteration $[{\tilde{z}}(k-1)\to {\tilde{z}}(k)]$ is 
 \[{\tilde{z}}(k)={-9 k^4\over -3+10k-10k^2+{\tilde{z}}(k-1)}\] 
  with
 asymptotics 
\[{\tilde{z}}(k)\simeq	{- 9k^4\over -10k^2+{\tilde{z}}(k)}\Rightarrow \lim_{k\to\infty}{{\tilde{z}}(k)\over k^2}=\{ 1, 9\}\] 
here $9k^2$ since one iterates from the Bessel integrals initial condition ${\tilde{z}}(1)$.

Using (\ref{rec}) the $k=1$ ($n=2$) initial conditions are reexpressed  on the   basis (\ref{ba2bis}) for $\kappa=3$, i.e. on $\{1,\quad I_{0,0}^{(3)}\}$, as 
\be I_{2,0}^{(3)}=-{1\over 3}\times 1+{2\over 3}\times I_{0,0}^{(3)}\quad\quad  
I_{2,2}^{(3)}= {1\over 6}\times 1 \Rightarrow u(1)\equiv {I_{2,2}^{(3)}\over I_{2,0}^{(3)}}={1\over -2+4 I_{0,0}^{(3)}} \label{originalter} \ee

\[\Rightarrow z(1)=-{-23 + 40 I_{0,0}^{(3)}\over -1 + 2 I_{0,0}^{(3)}}\Rightarrow z(0)=-3 + {3\over 2 I_{0,0}^{(3)}}\]
where again a simplification has occured as in (\ref{simple}).
% since one would have expected  $z(0)={a+b\;I_{0,0}^{(3)}\over c+d\;I_{0,0}^{(3)}}$.

So the rational approximation
\be\label{rat3}z(0)=-3 + {3\over 2 I_{0,0}^{(3)}}=\lim_{k\to\infty}{{y(k)}_{\{y(0)= 1,y(1)= 0\}}\over y(k)_{\{y(0)= 0,y(1)= 1\}}}\ee
with
\be\label{fractionbister}y(k+1)-(10k^2+10k+3)y(k)+9k^4y(k-1)=0\rightarrow y^2-10 y+9=0\rightarrow \{9,\; 1\}\ee
 
 The polynomial  $10k^2+10k+3=11k^2+11k+3-k(k+1)$ in the denominator of  (\ref{rec3}) as well as  $k^4$ in the numerator (with the opposite  sign) are again "Ap\'ery like". Indeed   Ap\'ery's proof of $\zeta(2)$ irrationality\footnote{There are  two other cases (see van der Poorten's review \cite{VanPoorten})  
 
  $z(k-1)={\displaystyle+8k^4\over \displaystyle({7k^2+7k}+2)+z(k)}\Rightarrow
  z(0)= {4\over  \zeta(2)}-2$
  
  and 
  
  $z(k-1)={\displaystyle+k^4(4k+1)(4k-1)\over \displaystyle{(2k+1)(3k^2+3k}+{\bf 1})+z(k)}\Rightarrow
  z(0)= {5\over2  \zeta(2)}-{\bf 1}$
  
  We found another case of the last type by a PSLQ search, see Section (\ref{sum}) below.} reads
  $$z(k-1)={+k^4\over 11k^2+11k+3+z(k)}$$
  $$z(0)= {5\over  \zeta(2)}-3$$
This is not a coincidence: for weight $\kappa=3$  it is possible to integrate  $I_{0,0}^{(3)}=\int_0^{\infty}uK_0(u)^3\,\rmd u $ to be proportionnal to a  $\zeta(2)$-like number\footnote{See Appendix A.}
  \[\int_0^{\infty}uK_0(u)^3\,\rmd u={\psi_1(1/3)-\psi_1(2/3)\over 12} \]
keeping in mind that  
  \[\zeta(2)={\psi_1(1/3)+\psi_1(2/3)\over 8}\]
 where $\psi_n(z)$ is the PolyGamma function
\be \psi_n(z)	=	(-1)^{n+1}n!\sum_{k=0}^{\infty}{1\over(z+k)^{n+1}}\ee 
  Again the numerical convergence of (\ref{rat3}) does not seem   sufficiently fast to conclude to the irrationality  of $z(0)$, that is of $I_{0,0}^{(3)}$,  that is of $\psi_1(1/3)-\psi_1(2/3)$,  whose  irrationality is unknown.

\subsection{Continuous fractions summary}\label{sum}
 \begin{itemize}
\item{\bf $\zeta(2)$: Ap\'ery, two other cases in Van der Poorten, PSLQ numerical}
 
  $z(k-1)={\displaystyle+k^4\over\displaystyle (11k^2+11k+{\bf 3})+z(k)}\Rightarrow z(0)={5\over  \zeta(2)}-{\bf 3}$
  
  $z(k-1)={\displaystyle+8k^4\over \displaystyle({7k^2+7k}+{\bf 2})+z(k)}\Rightarrow
  z(0)= {4\over  \zeta(2)}-{\bf 2}$
  
% $z(k-1)={\displaystyle+16k^6\over \displaystyle{(2k+1)(3k^2+3k}+{\bf 1})+z(k)}-{\displaystyle k^4\over \displaystyle{(2k+1)(3k^2+3k}+{\bf %1})+z(k)}\Rightarrow
%  z(0)= {5\over2  \zeta(2)}-{\bf 1}$

%$z(k-1)={\displaystyle+27k^6\over \displaystyle{(2k+1)(13k^2+13k}+{\bf 4})+z(k)}-{\displaystyle 3k^4\over \displaystyle{(2k+1)(13k^2+13k}+{\bf %4})+z(k)}\Rightarrow
%  z(0)= {7\over  \zeta(2)}-{\bf 4}$

  $z(k-1)={\displaystyle+k^4(4k+1)(4k-1)\over \displaystyle{(2k+1)(3k^2+3k}+{\bf 1})+z(k)}\Rightarrow
  z(0)= {5\over2  \zeta(2)}-{\bf 1}$

$z(k-1)={\displaystyle+3k^4(3k+1)(3k-1)\over \displaystyle{(2k+1)(13k^2+13k}+{\bf 4})+z(k)}\Rightarrow
  z(0)=_{\rm PSLQ} {7\over  \zeta(2)}-{\bf 4}$

\item{\bf  $  \psi_1(1/3)-\psi_1(2/3): \kappa=3 $}
 
 $z(k-1)={\displaystyle-9k^4\over \displaystyle (10k^2+10k+{\bf 3})+z(k)}\Rightarrow z(0)={18\over \psi_1(1/3)-\psi_1(2/3)}-{\bf 3} $
 
 %$z(k-1)={\displaystyle-36k^6\over \displaystyle{(2k+1)(10k^2+10k}+{\bf 3})+z(k)}+{\displaystyle 9k^4\over \displaystyle{(2k+1)(10k^2+10k}+{\bf %3})+z(k)}\Rightarrow
 % z(0)= {18\over \psi_1(1/3)-\psi_1(2/3)}-{\bf 3}$

\item{\bf $\zeta(3)$: Ap\'ery
%\footnote{Asymptotics : $k\to\infty$ one has $z(\infty)={\displaystyle- k^6\over\displaystyle 34 k^3+z(\infty)}\to 
%z(\infty)^2+ 34 k^3z(\infty)+k^6=0\to z(\infty)=-(1+\sqrt{2})^4k^3$ or $z(\infty)=-{1\over(1+\sqrt{2})^{4}}k^3$}
, PSLQ numerical, $\kappa=4$}
 
$z(k-1)={\displaystyle- k^6\over\displaystyle (2k+1)(17k^2+17k+{\bf 5})+z(k)}\Rightarrow z(0)={6\over \zeta(3)}-{\bf 5}$
 
$z(k-1)={\displaystyle- k^6\over \displaystyle(2k+1)(3k^2+3k+{\bf 1})+z(k)}\Rightarrow z(0)=_{\rm PSLQ}{ 8\over 7\zeta(3)}-{\bf 1}$

$z(k-1)={\displaystyle-4 k^6\over \displaystyle(2k+1)({5k^2+5k\over 2}+{\bf 1})+z(k)}\Rightarrow z(0)={6 \over 7\zeta(3)}-{\bf 1}$
\end{itemize}
{where one  notes}  
\begin{itemize}

\item the usual   $p k^2+p k +{\bf q}$  in the denominators
%\footnote{ If one rewrites the last $\zeta(3)$ continuous fraction as 
%$$z(k-1)={\displaystyle-k^6\over \displaystyle(2k+1)({5k^2+5k\over 4}+{\bf {1\over 2}})+z(k)}\Rightarrow z(0)={3 \over 7\zeta(3)}-{\bf {1\over %2}}$$ 
%then the  three $\zeta(3)$ continuous fractions  are such that    $p={7 q-1\over 2}$.} 
and {\bf   q} in the $z(0)$'s

\item the sign change\footnote{See also (\ref{sign}): from this point of view Ap\'ery's ${6\over \zeta(3)}-{ 5}$ could rather rewrite as  $ {42\over 7\zeta(3)}-{ 5}$.}   $+k^4\to -k^4$ in relation with  $\psi_1(1/3)+\psi_1(2/3)\to \psi_1(1/3)-\psi_1(2/3)$.

\end{itemize}

Various PSLQ searches  did not produce so far  any other example of continuous fractions of this type for $\zeta(2)$, $\zeta(3)$ or $\psi_1(1/3)-\psi_1(2/3)$ (for which only the case\footnote{This continuous fraction appears as a "sporadic" case in the numerical search \cite{Zagier} where it is related to  the Dirichlet L-series $L_3(2)$  which "is, up to a factor $3^{3/2}/4$, the maximum volume of a tetrahedron in hyperbolic 3-space".} listed above seems to be known).
%This might indicate that quadratic or more complicated numbers might play a role in any generalisation yet to be found.

\section{ TRYING TO INTEGRATE $\int_{0}^{\infty}\rmd u\,  u
K_{0}(u)^n$}\label{integrating}

%Consider $$\int_{0}^{\infty}\rmd u\,  u
%K_{0}(u)^n\quad n\ge 0$$
\small{As alluded to in Section (\ref{arxiv})  Feyman diagrams  momenta integrations  
 lead to multiple integrals on  differences of consecutive intermediate temperatures. Bessel integrals $\int_{0}^{\infty}\rmd u\,  u
K_{0}(u)^n$ can quite generally be represented in this particular way: 
change  variable $u=2\sqrt{t}$ and use the integral representation 
 \be\label{represent} \int_0^{\infty}\rmd a\,   a^{\nu -1}e^{ -a-{t\over a}}
=2K_{\nu}(u)({u\over 2})^{\nu}\ee
so that
\begin{align}&\int_{0}^{\infty}\rmd u\,  u
K_{0}(u)^n\nonumber\\
=&\int_{0}^{\infty}2\rmd t   \int_{0}^{\infty}{\rmd a_1\over 2} \ldots \int_{0}^{\infty}{\rmd a_n\over 2}{1\over a_1 a_2\ldots a_n}e^{ -a_1-{t\over a_1}}e^{ -a_2-{t\over a_2}}\ldots e^{ -a_n-{t\over a_n}}\nonumber\\
=&2  \int_{0}^{\infty}{\rmd a_1\over 2} \ldots \int_{0}^{\infty}{\rmd a_n\over 2}{1\over a_1 a_2\ldots a_n}{1\over{1\over a_1}+{1\over a_2}+\ldots+{1\over a_n}}e^{ -a_1-a_2-\ldots-a_n}\nonumber\\
=&{1\over 2^{n-1}}  \int_{0}^{\infty}{\rmd a_1} \ldots \int_{0}^{\infty}{\rmd a_n}{1\over a_1 a_2\ldots a_{n-1}+ a_2 a_3\ldots a_{n}+\ldots+ a_{n} a_1\ldots a_{n-2}}e^{ -(a_1+a_2+\ldots+a_n)}\label{feyn}\end{align}

{\bf $a_{n}$ integration:} %change variables $a_i\to a'_i={ a_i/ s}$ (notation $a'_i\to a_i$)
%\begin{align}&\int_{0}^{\infty}\rmd u\,  u
%K_{0}(u)^n\nonumber\\&={s\over 2^{n-1}}  \int_{0}^{\infty}{\rmd a_1} \int_{0}^{\infty}{\rmd a_2}\ldots \int_{0}^{\infty}{\rmd a_n}
%{1\over a_1 a_2\ldots a_{n-1}+ a_2 a_3\ldots a_{n}+\ldots+ a_{n} a_1\ldots a_{n-2}}e^{ -s(a_1+a_2+\ldots+a_n)}\end{align}
introduce the variable $\beta$ to rewrite $e^{ -(a_1+a_2+\ldots+a_n)}=\int_{0}^{\infty}{\rmd} \beta\,e^{-\beta}\delta(\beta-(a_1+a_2+\ldots+a_n))$, then integrate over $a_n$
\begin{align}&\int_{0}^{\infty}\rmd u\,  u
K_{0}(u)^n\nonumber\\   
%=&{1\over 2^{n-1}} \int_{0}^{\infty}{\rmd \beta} e^{-\beta } \int_{0}^{\infty}{\rmd a_1} \ldots \int_{0}^{\infty}{\rmd a_n}{1\over a_1 a_2\ldots %a_{n-1}+ a_2 a_3\ldots a_{n}+\ldots+ a_{n} a_1\ldots a_{n-2}}
%\delta(\beta-(a_1+\ldots+a_n))\nonumber\\
=&{1\over 2^{n-1}} \int_{0}^{\infty}{\rmd \beta} e^{-\beta } \int_{0}^{\beta}{\rmd a_1} \int_{0}^{\beta-a_1}{\rmd a_2}\ldots \int_{0}^{\beta-a_1-\ldots-a_{n-2}}{\rmd a_{n-1}}\nonumber\\ 
&{1\over a_1 a_2\ldots a_{n-1}+ (\beta-(a_1+a_2+\ldots+a_{n-1}))(a_2a_3 \ldots a_{n-1}+a_3a_4 \ldots a_{n-1}a_1+\ldots+  a_1\ldots a_{n-2} )}\end{align}
Change variables $a_i\to a'_i={ a_i/ \beta}$ (notation $a'_i\to a_i$)
%\begin{align}&\int_{0}^{\infty}\rmd u\,  u
%K_{0}(u)^n={1\over 2^{n-1}} \int_{0}^{\infty}{\rmd \beta} e^{-\beta } \int_{0}^{1}{\rmd a_1} \int_{0}^{1-a_1}{\rmd a_2}\ldots %\int_{0}^{1-%a_1-\ldots-a_{n-2}}{\rmd a_{n-1}}\nonumber\\ 
%&{1\over a_1 a_2\ldots a_{n-1}+ (1-(a_1+a_2+\ldots+a_{n-1}))(a_2a_3 \ldots a_{n-1}+a_3a_4 \ldots a_{n-1}a_1+\ldots+  a_1\ldots a_{n-2}  %)}\end{align}
{so that the $\beta$ integration becomes trivial and finally}  
\begin{align}&\int_{0}^{\infty}\rmd u\,  u
K_{0}(u)^n={1\over 2^{n-1}} \int_{0}^{1}{\rmd a_1} \int_{0}^{1-a_1}{\rmd a_2}\ldots \int_{0}^{1-a_1-\ldots-a_{n-2}}{\rmd a_{n-1}}\nonumber\\ 
&{1\over a_1 a_2\ldots a_{n-1}+ (1-(a_1+a_2+\ldots+a_{n-1}))(a_2a_3 \ldots a_{n-1}+a_3a_4 \ldots a_{n-1}a_1+\ldots+  a_1a_2\ldots a_{n-2} )}\end{align}
%$${1\over a_1 a_2\ldots a_{n-2}(1- (a_1 +a_2+\ldots+ a_{n-2}))+ a_{n-1}(1-(a_1+a_2+\ldots+a_{n-2})-a_{n-1})(a_2a_3 \ldots a_{n-2}+a_3a_4 \ldots a_{n-2}a_1+\ldots+ a_{n-2}a_1 a_2\ldots a_{n-4}+a_1a_2\ldots a_{n-3} )}$$
With  the notations  
 
$u_{n}=a_1 +a_2+\ldots+ a_{n}$ 

$v_{n}= a_2a_3 \ldots a_{n}+a_3a_4 \ldots a_{n}a_1+\ldots+ a_{n}a_1 a_2\ldots a_{n-2}+a_1a_2\ldots a_{n-1}$ 

$w_{n}=a_1 a_2\ldots a_{n}$

one has shown that
\begin{align}\int_{0}^{\infty}\rmd u\,  u
K_{0}(u)^n&={1\over 2^{n-1}}  \int_{0}^{\infty}{\rmd a_1} \int_{0}^{\infty}{\rmd a_2}\ldots \int_{0}^{\infty}{\rmd a_n}{1\over v_n}e^{ -u_n}\nonumber\\
&={1\over 2^{n-1}} \int_{0}^{1}{\rmd a_1} \int_{0}^{1-u_1}{\rmd a_2}\ldots \int_{0}^{1-u_{n-2}}{\rmd a_{n-1}}{1\over w_{n-1}+ (1-u_{n-1}) v_{n-1}}\label{etoui}
\end{align}
{\bf $a_{n-1}$ integration:}
use 
$$ u_{n-1}=a_{n-1}+u_{n-2}, \quad v_{n-1}=a_{n-1}v_{n-2}+w_{n-2}, \quad w_{n-1}=a_{n-1}w_{n-2}$$
so that
\begin{align}
&{1\over w_{n-1}+ (1-u_{n-1}) v_{n-1}}
= {1\over w_{n-2}(1-u_{n-2})+a_{n-1}v_{n-2}(1-u_{n-2})-a_{n-1}^2 v_{n-2}}\nonumber\\
&={1\over -v_{n-2}(a_{n-1}-a_{n-1}^+)(a_{n-1}-a_{n-1}^-)} ={1\over- v_{n-2}(a_{n-1}^{+}-a_{n-1}^{-})}({1\over a_{n-1}-a_{n-1}^{+}}-{1\over a_{n-1}-a_{n-1}^{-} })\end{align}
 
where $$a_{n-1}^{\pm}={-v_{n-2}(1-u_{n-2})\pm\sqrt{v_{n-2}^2(1-u_{n-2})^2+4w_{n-2}v_{n-2}(1-u_{n-2})}\over -2v_{n-2}}$$

One finds
\begin{align} 
&\int_{0}^{\infty}\rmd u\,  u
K_{0}(u)^n={1\over 2^{n-1}} \int_{0}^{1}{\rmd a_1} \int_{0}^{1-u_1}{\rmd a_2}\ldots \int_{0}^{1-u_{n-3}}{\rmd a_{n-2}}\nonumber\\ 
&{1\over \sqrt{v_{n-2}^2(1-u_{n-2})^2+4w_{n-2}v_{n-2}(1-u_{n-2})}}\log|{a_{n-1}^{-}\over a_{n-1}^{+}}{1-u_{n-2}-a_{n-1}^{+}\over 1-u_{n-2}-a_{n-1}^{-}}|\end{align}
where the integrand rewrites as
\begin{align}{2\over \sqrt{v_{n-2}^2(1-u_{n-2})^2+4w_{n-2}v_{n-2}(1-u_{n-2})}}\log|{a_{n-1}^{-}\over a_{n-1}^{+}}|={2\over (1-u_{n-2})v_{n-2}}X\log{1+X\over 1-X}\end{align}
with $$X=\sqrt{1\over 1+{4w_{n-2}\over (1-u_{n-2})v_{n-2}}}$$
Change variables $a_i\to x_i=X a_i/u_{n-2}$ so that when $0<a_i<1$ then  $1>X>0\Rightarrow 1>x_i>0$. The    Jacobian  is
$$ {(u_{n-2})^{n-2}\over X^{n-3}|\sum_{i=1}^{n-2}a_i\partial_i X|}$$
Use homogeneity relations
\begin{align}
\sum_{i=1}^{n-2}a_i\partial_i u_{n-2}&=u_{n-2}\nonumber\\
\sum_{i=1}^{n-2}a_i\partial_i v_{n-2}&=(n-3)v_{n-2}\nonumber\\ 
\sum_{i=1}^{n-2}a_i\partial_i w_{n-2}&=(n-2)w_{n-2}\end{align}
so that
%$$\sum_{i=1}^{n-2}a_i\partial_i X={-4v_{n-2} w_{n-2}\over 2 X((1-u_{n-2})v_{n-2}+4w_{n-2})^2}$$
$$\sum_{i=1}^{n-2}a_i\partial_i X=-{1\over 2}X(1-X^2){1\over 1-u_{n-2}}$$
and
\begin{align}\int_{0}^{\infty}\rmd u\,  u
K_{0}(u)^n=&{1\over 2^{n-1}} \int_{0}^{1}{\rmd x_1} \int_{0}^{1-x_1}{\rmd x_2}\ldots \int_{0}^{1-x_1-\ldots - x_{n-3}}{\rmd x_{n-2}}\log{1+X\over 1-X}\nonumber\\
&{2\over (1-u_{n-2})v_{n-2}}X{(u_{n-2})^{n-2}\over X^{n-3}|-{1\over 2}X(1-X^2){1\over 1-u_{n-2}}| }\nonumber\\
=&{1\over 2^{n-1}} \int_{0}^{1}{\rmd x_1} \int_{0}^{1-x_1}{\rmd x_2}\ldots \int_{0}^{1-x_1-\ldots - x_{n-3}}{\rmd x_{n-2}}\log{1+X\over 1-X}
{4\over v_{n-2}}{(u_{n-2})^{n-2}\over X^{n-3}(1-X^2) }\end{align} 
Use 
\begin{align}w_{n-2} X^{n-2}&=(u_{n-2})^{n-2} x_1x_2\ldots x_{n-2}\nonumber\\ 
{1\over X^2}-1&={4w_{n-2}\over(1-u_{n-2})v_{n-2}}\label{ouaip} \end{align}
%forget about the $||$ 
so that
$$\int_{0}^{\infty}\rmd u\,  u
K_{0}(u)^n={1\over 2^{n-1}} \int_{0}^{1}{\rmd x_1} \int_{0}^{1-x_1}{\rmd x_2}\ldots \int_{0}^{1-x_1-\ldots - x_{n-3}}{\rmd x_{n-2}}\log{1+X\over 1-X}{1\over X}{1-u_{n-2}\over  x_1x_2\ldots x_{n-2}}$$ 
Use  
\begin{align}X=&x_1+x_2+\ldots+x_{n-2}\nonumber\\
{u_{n-2}v_{n-2}\over w_{n-2}X}=&{1\over x_1}+{1\over x_2}+\ldots+{1\over x_{n-2}}\nonumber\end{align}
and (\ref{ouaip}) rewritten as
\be {4w_{n-2}\over (1-u_{n-2})v_{n-2}}={1\over (x_1+x_2+\ldots+x_{n-2})^2}-1\nonumber\ee
so that finally 
%by $a_{n-1}$ integration
\begin{align}\label{nicenice}&\int_{0}^{\infty}\rmd u\,  u
K_{0}(u)^n ={1\over 2^{n-1}} \int_{0}^{1}{\rmd x_1} \int_{0}^{1-x_1}{\rmd x_2}\ldots \int_{0}^{1-x_1-\ldots - x_{n-3}}{\rmd x_{n-2}}\log{1+x_1+x_2+\ldots+x_{n-2}\over 1-(x_1+x_2+\ldots+x_{n-2})}\nonumber\\
&{4\over 4 (x_1+\ldots+x_{n-2})x_1\ldots x_{n-2}+(1-(x_1+\ldots+x_{n-2})^2)(x_2x_3\ldots x_{n-2}+\ldots+x_1x_2\ldots x_{n-3})}
\end{align} 

This integration  generalises to $\int_{0}^{\infty}\rmd u\,  u^p
K_{0}(u)^n$ with $p$ odd: for example
 \begin{align}&\int_{0}^{\infty}\rmd u\,  u^3
K_{0}(u)^n={1\over 2^{n-3}} \int_{0}^{1}{\rmd a_1} \int_{0}^{1-a_1}{\rmd a_2}\ldots \int_{0}^{1-a_1-\ldots-a_{n-2}}{\rmd a_{n-1}}\nonumber\\
&{a_1a_2\ldots a_{n-1}(1- (a_1 +a_2+\ldots+ a_{n-1}))\over (a_1 a_2\ldots a_{n-1}+ (1-(a_1+a_2+\ldots+a_{n-1}))(a_2a_3 \ldots a_{n-1}+a_3a_4 \ldots a_{n-1}a_1+\ldots+  a_1a_2\ldots a_{n-2} ))^2}\nonumber\\
&={1\over 2^{n-3}} \int_{0}^{1}{\rmd x_1} \int_{0}^{1-x_1}{\rmd x_2}\ldots \int_{0}^{1-x_1-\ldots - x_{n-3}}{\rmd x_{n-2}}\nonumber\\
&\left({1+(x_1+x_2+\ldots+x_{n-2})^2\over 2(x_1+x_2+\ldots+x_{n-2})}\log{1+x_1+x_2+\ldots+x_{n-2}\over 1-(x_1+x_2+\ldots+x_{n-2})}-1\right)\nonumber\\ 
&{4x_1\ldots x_{n-2}\left(1-(x_1+x_2+\ldots+x_{n-2})^2\right)\over \bigg(4 (x_1+\ldots+x_{n-2})x_1\ldots x_{n-2}+(1-(x_1+\ldots+x_{n-2})^2)(x_2x_3\ldots x_{n-2}+\ldots+x_1x_2\ldots x_{n-3})\bigg)^2}\label{cela}\end{align} 

Finally if in (\ref{nicenice}) and (\ref{cela}) one changes notations $x_i\to a_i$  one has  shown
%$$ \int_{0}^{\infty}{\rmd a_1} \int_{0}^{\infty}{\rmd a_2}}\ldots \int_{0}^{\infty}{\rmd a_{n-1}}\int_{0}^{\infty}{\rmd a_{n}}\quad{1\over %a_2a_3 \ldots a_{n}+a_3a_4 \ldots a_{n}a_1+\ldots+  a_1a_2\ldots a_{n-1}}\quad e^{-a_1+a_2+\ldots+a_{n}}$$ 
%$$= \int_{0}^{1}{\rmd a_1} \int_{0}^{1-a_1}{\rmd a_2}\ldots \int_{0}^{1-a_1-\ldots - a_{n-3}}{\rmd a_{n-3}} \int_{0}^{1-a_1-\ldots-a_{n-2}}{\rmd %a_{n-1}}$$ 
%$${1\over a_1 a_2\ldots a_{n-1}+ (1-(a_1+a_2+\ldots+a_{n-1}))(a_2a_3 \ldots a_{n-1}+a_3a_4 \ldots a_{n-1}a_1+\ldots+  a_1a_2\ldots a_{n-2} )}$$ 
%$$=  \int_{0}^{1}{\rmd a_1}\int_{0}^{1-a_1}{\rmd a_2}\ldots \int_{0}^{1-a_1-\ldots - a_{n-3}}{\rmd %a_{n-2}}\quad\log{1+a_1+a_2+\ldots+a_{n-2}\over 1-(a_1+a_2+\ldots+a_{n-2})}$$ 
%$${4\over 4 (a_1+a_2+\ldots+a_{n-2})a_1a_2\ldots a_{n-2}+(1-(a_1+a_2+\ldots+a_{n-2})^2)(a_2a_3\ldots a_{n-2}+\ldots+a_1a_2\ldots a_{n-3})}$$ 
%
%one has  shown 
\begin{align} &2^{n-1}\int_{0}^{\infty}{\rmd u}\;u K_0(u)^n=\int_{0}^{\infty}{\rmd a_1}\ldots \int_{0}^{\infty}{\rmd a_{n-2}}\int_{0}^{\infty}{\rmd a_{n-1}}\int_{0}^{\infty}{\rmd a_{n}}\quad{1\over v_n}\quad e^{-u_{n}}\nonumber\\
&= \int_{0}^{1}{\rmd a_1} \int_{0}^{1-u_1}{\rmd a_2}\ldots \int_{0}^{1-u_{n-3}}{\rmd a_{n-2}}\int_{0}^{1-u_{n-2}}{\rmd a_{n-1}}\quad{1\over w_{n-1}+ (1-u_{n-1})v_{n-1}}\label{OK}\\
&=  \int_{0}^{1}{\rmd a_1}\int_{0}^{1-u_1}{\rmd a_2}\ldots \int_{0}^{1-u_{n-3}}{\rmd a_{n-2}}\quad\log{1+u_{n-2}\over 1-u_{n-2}}\quad{4\over 4 u_{n-2}w_{n-2}+(1-u_{n-2}^2)v_{n-2}}\label{jean}\end{align}
%the last equality does not seem to imply the simple integral
%$$  \int_{0}^{1-u_{n-2}}{\rmd a_{n-1}}\quad{1\over w_{n-1}+ (1-u_{n-1})v_{n-1}}= \quad \log{1+u_{n-2}\over 1-u_{n-2}}\quad{4\over 4 %u_{n-2}w_{n-2}+(1-u_{n-2}^2)v_{n-2}}$$ 
%$${1\over a_1 a_2\ldots a_{n-2}(1- (a_1 +a_2+\ldots+ a_{n-2}))+ a_{n-1}(1-(a_1+a_2+\ldots+a_{n-2})-a_{n-1})(a_2a_3 \ldots a_{n-2}+a_3a_4 \ldots a_{n-2}a_1+\ldots+ a_{n-2}a_1 a_2\ldots a_{n-4}+a_1a_2\ldots a_{n-3} )}$$
  %$\tilde\zeta(3)=7\zeta(3)/2$.

%so that   the $a_{n-1}$ integration amounts to
%$${1\over w_{n-1}+ (1-u_{n-1})v_{n-1}}\to\log{1+u_{n-2}\over 1-u_{n-2}}\quad{4\over 4 u_{n-2}w_{n-2}+(1-u_{n-2}^2)v_{n-2}}$$ 
and
\begin{align}&2^{n-3}\int_{0}^{\infty}{\rmd u}\;u^3 K_0(u)^n= \int_{0}^{\infty}{\rmd a_1}\ldots \int_{0}^{\infty}{\rmd a_{n-2}}\int_{0}^{\infty}{\rmd a_{n-1}}\int_{0}^{\infty}{\rmd a_{n}}\quad{w_n\over v_n^2}\quad e^{-u_{n}}\nonumber\\ 
&= \int_{0}^{1}{\rmd a_1} \int_{0}^{1-u_1}{\rmd a_2}\ldots \int_{0}^{1-u_{n-3}}{\rmd a_{n-2}}\int_{0}^{1-u_{n-2}}{\rmd a_{n-1}}\quad{w_{n-1}(1-u_{n-1})\over (w_{n-1}+ (1-u_{n-1})v_{n-1})^2}\label{OKK}\\ 
&= \int_{0}^{1}{\rmd a_1}\int_{0}^{1-u_1}{\rmd a_2}\ldots \int_{0}^{1-u_{n-3}}{\rmd a_{n-2}}\left({1+u_{n-2}^2\over2u_{n-2} }\log{1+u_{n-2}\over 1-u_{n-2}}-1\right){4w_{n-2}(1-u_{n-2}^2)\over( 4 u_{n-2}w_{n-2}+(1-u_{n-2}^2)v_{n-2})^2}\nonumber\end{align}
%so that the $a_{n-1}$ integration amounts to,
%$${w_{n-1}(1-u_{n-1})\over (w_{n-1}+ (1-u_{n-1})v_{n-1})^2}\to\left({1+u_{n-2}^2\over2u_{n-2} }\log{1+u_{n-2}\over 1-%u_{n-2}}-1\right)\quad{4w_{n-2}(1-u_{n-2}^2)\over( 4 u_{n-2}w_{n-2}+(1-u_{n-2}^2)v_{n-2})^2}$$ 
(\ref{OK}) and (\ref{OKK}) indicate that $\int_{0}^{\infty}{\rmd u}\;u K_0(u)^n$ and $\int_{0}^{\infty}{\rmd u}\;u^3 K_0(u)^n$ are periods.
This is the case in general for  $\int_{0}^{\infty}{\rmd u}\;u^m K_0(u)^n$ with $m$ odd.

}

\section{CONCLUSION : WHY LOOKING AT BESSEL INTEGRALS ? IS $\zeta(5)$ IRRATIONAL ?}

In Section (\ref{arxiv}) some arguments (and examples) were given of why weight $\kappa$ Bessel integrals are expected to be related to weight $\kappa-1$ (and below)  zeta numbers.  
%One  knows for example that

 %\[{\psi_1(1/3)-\psi_1(2/3)=12\int_0^{\infty}\rmd u\,  u
%K_{0}(u)^3}\]%=12 I_{0,0}^{(3)}\]
%and
  %\[{\zeta(3)}={8\over 7}\int_0^{\infty}\rmd u\,  u
%K_{0}(u)^4\]%={8\over 7}I_{0,0}^{(4)}\]
One  has found  the surprisingly simple PSLQ identity
\be\label{final} \zeta(5)=_{\rm PSLQ}{1\over 77}\int_{0}^{\infty}u \, K_0(u)^8\,\rmd u
-{72\over 77}\int_{0}^{\infty}u^3 \, K_0(u)^8\,\rmd u %={1\over 77}I_{0,0}^{(8)}-{144\over 77}I_{2,0}^{(8)}
\ee  
 If one could 
 prove  for $\kappa=8$ the irrationality claim on (\ref{ba1bis}), in particular on    $\int_{0}^{\infty}u \, K_0(u)^8\,\rmd u$ and $\int_{0}^{\infty}u^3 \, K_0(u)^8\,\rmd u$,
 and  derive\footnote{Note added on Monday, November 16, 2026 : (\ref{final}) has just been proved in https://zenodo.org/records/24247590.}
  (\ref{final}),
   then one might have a proof of the irrationality of $\zeta(5)$. 
   A related issue is   to generalize the algebraic construction of Section (\ref{goodsection}) to weights $\kappa\ge 5$,  with  Bessel basis of dimension $\ge  3$ (see  Appendix B).
   
Coming back in Section (\ref{sum})  to $\zeta(3)$, $\zeta(2)$ and $\psi_1(1/3)-\psi_1(2/3)$, it would be rewarding to find a systematics  beyond the cases listed 
%The overall presence of the hallmark Ap\'ery polynomial  $pk^2+pk+q$ in $z(k)$  and  of $q$ in $z(0)$ are  possible indications of  a more %general structure  yet to be found. 
and have simple expressions (Ap\'ery-like numbers) for the $y_n$'s satisfying the recurrence relations associated to these continuous fractions  with initial conditions $\{ y(0)=1,y(1)=0\}$  or $\{ y(0)=0,y(1)=1\}$.  With the Ap\'ery initial conditions  $\{y(0)=1,y(0)=q\}$, the recurrence
$y(k+1)-(2k+1)(3k^2+3k+{1})y(k)+k^6y(k-1)=0$ associated to the continuous fraction
 \be z(k-1)={\displaystyle- k^6\over \displaystyle(2k+1)(3k^2+3k+{\bf 1})+z(k)}\Rightarrow z(0)=_{\rm PSLQ}{ 8\over 7\zeta(3)}-{\bf 1}\nonumber\ee
  has for solution \cite{Zu} 
\be \label{zu1}y(k)_{\{y(0)=1,y(1)=1\}}={k!^3\over 2^{2k}}\sum_{i=0}^k \left({k\atop i}\right)^2 \left({2 i\atop k}\right)^2\ee
where $k!^3$ factorises. 
Similarly for  the  $\kappa=4$ continuous 
fraction 
\be z(k-1)={\displaystyle -16 k^6\over \displaystyle(2k+1)({5k^2+5k}+{\bf 2})+z(k)}\Rightarrow z(0)={12 \over 7\zeta(3)}-{\bf 2}\ee
\be \label{zu2}y(k)_{\{y(0)=1,y(1)=2\}}={k!^3\over 2^{k}}\sum_{i=0}^k \left({k\atop i}\right)^2 \left({2 i\atop i}\right) \left({2 (k-i)\atop k-i}\right)\ee
(see in \cite{Zagier} the  solutions for the second $\zeta(2)$  and the $\kappa=3$  continuous fractions listed in Section (\ref{sum})). 

One would also like to   integrate  $\int_0^{\infty}u\,K_0(u)^n\,\rmd u$ further up to possibly one integration  left on an integrand which should contain a term of the type $(\log(1+x)/(1-x))^{n-2}$ times a rational function of $x$ yet to be determined, possibly allowing for a derivation of (\ref{final}).

Finally the fact that Bessel integrals fall in the category of periods might also be an indication of a deeper meaning yet to be  understood.
  
  Acknowledgments: S.O. acknowledges some useful conversations with  S. Mashkevich in particular for helping in the numerics involved in the PSLQ searches of Appendix C. We also would like to thank Alain Comtet  for a careful reading of the manuscript. 
\pagebreak
%\vspace=5cm
%  and if one notes $z(0)+q={r\over I_{0,0}^{(4)}} $ then $16r= {50q\over 3}   + 1 - {5\over 3  q}$.

\section*{ APPENDIX A: INTEGRATING FURTHER  }

\subsection*{1) Trying to integrate $\int_{0}^{\infty}\rmd u\,  u
I_0(u)K_{0}(u)^n$}

Using again the integral representation (\ref{represent}) and ($u=2\sqrt{t}$)
\be I_0(u)=\sum_{k=0}^{\infty}{t^k\over (k!)^2}\ee
one can integrate over $t$ using 
\be \int_0^{\infty} \rmd t\; t^ke^{-t x}={k!\over x^{k+1}}\ee
to obtain 
\begin{align} &2^{n-1}\int_{0}^{\infty}{\rmd u}\;u I_0(u)K_0(u)^n=\int_{0}^{\infty}{\rmd a_1}\ldots \int_{0}^{\infty}{\rmd a_{n-2}}\int_{0}^{\infty}{\rmd a_{n-1}}\int_{0}^{\infty}{\rmd a_{n}}\quad{1\over v_n}\quad e^{\displaystyle -u_{n}+{w_n\over v_n}}\nonumber
\end{align}
Introducing as above the variable $\beta$, integrating over $a_n$ and then trivially over $\beta$ one finally obtains
\begin{align} &2^{n-1}\int_{0}^{\infty}{\rmd u}\;u I_0(u) K_0(u)^n\nonumber\\
&= \int_{0}^{1}{\rmd a_1} \int_{0}^{1-u_1}{\rmd a_2}\ldots \int_{0}^{1-u_{n-3}}{\rmd a_{n-2}}\int_{0}^{1-u_{n-2}}{\rmd a_{n-1}}\quad{1\over w_{n-1}u_{n-1}+ (1-u_{n-1})v_{n-1}}\label{OKKK}\end{align}
a result  to be compared to (\ref{etoui}). 

One can push the integration one step further following the same procedure as in Section (\ref{integrating}) to obtain an  expression again in terms of $u_{n-2}, v_{n-2}$ and $w_{n-2}$ but somehow  more involved than (\ref{jean}) since its denominator  contains a square root.

\subsection*{2) More on integrating  $\int_{0}^{\infty}\rmd u\,  u
K_{0}(u)^n$, $\int_{0}^{\infty}\rmd u\,  u^3 K_{0}(u)^n$ and $\int_{0}^{\infty}\rmd u\,  u
I_0(u)K_{0}(u)^n$
}
\small{

{$\bf n=3$}

$$\int_{0}^{\infty}\rmd u\,  u
K_{0}(u)^3={1\over 2^{2}} \int_{0}^{1}{\rmd x_1}\log{1+x_1\over 1-x_1} 
{4\over 4 (x_1)x_1+(1-(x_1)^2)}=\int_{0}^{1}{\rmd x_1}\log{1+x_1\over 1-x_1} 
{1\over 1+3x_1^2}$$ 

{ $\bf n=4$}

\begin{align}\int_{0}^{\infty}\rmd u\,  u
K_{0}(u)^4&={1\over 2^{3}} \int_{0}^{1}{\rmd x_1}\int_{0}^{1-x_1}{\rmd x_2}\log{1+x_1+x_2\over 1-x_1-x_2} {1\over x_1+x_2} 
{4\over 4 x_1x_2+(1-(x_1+x_2)^2)}\end{align}

{$\bf n=5$}
\begin{align}\int_{0}^{\infty}\rmd u\,  u
K_{0}(u)^5=&{1\over 2^{4}} \int_{0}^{1}{\rmd x_1} \int_{0}^{1-x_1}{\rmd x_2} \int_{0}^{1-x_1- x_{2}}{\rmd x_{3}}\log{1+x_1+x_2+x_{3}\over 1-(x_1+x_2+x_{3})}\nonumber\\ 
&{4\over 4 (x_1+x_2+x_3)x_1x_2x_3+(1-(x_1+x_2+x_3)^2)(x_2x_3+x_3x_1+x_1x_2)}\end{align} 

Integrating further

{$\bf n=3$}

 %\begin{align}\int_{0}^{\infty}\rmd u\,  u
%K_{0}(u)^3&= \int_{0}^{1}{\rmd x_1}\log{1+x_1\over 1-x_1} 
%{1\over 1+3x_1^2}={2} \int_{0}^{1}{\rmd x_1}\sum_{n=0}^{\infty}{x_1^{2n+1}\over 2n+1}{1\over 1+3x_1^2}\nonumber\\ 
%=\sum_{0}^{\infty}{1\over 2n+1}\int_{0}^{1}{\rmd x_1}{x_1^{n}\over 1+3x_1}
%&=\sum_{0}^{\infty}{1\over 2n+1}\int_{0}^{1}{\rmd x_1}{x_1^{n}\over 1+3x_1}={\psi_1(1/3)-\psi_1(2/3)\over 12}\end{align}
%where $[n,z]$ is the PolyGamma 
%$$ [n,z]\equiv \psi_n(z)	=	(-1)^{n+1}n!\sum_{k=0}^{\infty}{1\over(z+k)^{n+1}} $$
%or rather, 
 \begin{align}\int_{0}^{\infty}\rmd u\,  u
K_{0}(u)^3&= \int_{0}^{1}{\rmd x_1}\log{1+x_1\over 1-x_1} 
{1\over 1+3x_1^2}= \int_{0}^{1}{\rmd x_1}\log{1+x_1\over 1-x_1} 
{2\over (1+x_1)^3(1+({1-x_1\over 1+x_1})^3)}\nonumber\\
&=-{1\over 2}
\int_{0}^{1}{\rmd u}{1+u\over 1+u^3}\log{u}\nonumber\\
&=-{1\over 2}
\int_{0}^{1}{\rmd u}(1+u)\sum_{k=0}^{\infty}(-1)^ku^{3k}\log{u} \nonumber\\
&={1\over 2}\sum_{k=0}^{\infty}(-1)^k({1\over(3k+1)^2}+{1\over(3k+2)^2})={\psi_1(1/3)-\psi_1(2/3)\over 12} 
\end{align}
where one has made the change of variable $u={1-x_1\over 1+x_1}$ and used \be\int_0^1\rmd u\, u^k(\log u)^n =(-1)^n{n!\over (k+1)^n}\ee 

%\begin{align}\int_{0}^{\infty}\rmd u\,  u^3
%K_{0}(u)^3&= \int_{0}^{1}{\rmd x_1}\left({1+x_1^2\over 2x_1}\log{1+x_1\over 1-x_1}-1\right) 
%{4x_1(1-x_1^2)\over (1+3x_1^2)^2}
%\end{align}
{$\bf n=4$}
\begin{align}\int_{0}^{\infty}\rmd u\,  u
K_{0}(u)^4&={1\over 2^{3}} \int_{0}^{1}{\rmd x_1}\int_{0}^{1-x_1}{\rmd x_2}\log{1+x_1+x_2\over 1-x_1-x_2} {1\over x_1+x_2}
{4\over 1-(x_1+x_2)^2+4 x_1x_2}\nonumber\\ 
&={1\over 2^{3}}{1\over 2} \int_{0}^{1}{\rmd x}{1\over x}\log{1+x\over 1-x}\int_{-x}^{x}{\rmd y} 
{4\over 1-y^2}\nonumber\\
&={1\over 2} \int_{0}^{1}{\rmd x}{1\over x}\log{1+x\over 1-x}\int_0^{x}{\rmd y} 
{1\over 1-y^2}\nonumber\\
&={1\over 2}{1\over 2} \int_{0}^{1}{\rmd x}{1\over x}(\log{1+x\over 1-x})^2 ={7\zeta(3)\over 8}\label{cb}
\end{align}
where one has made the change of variables $x=x_1+x_2$ and $y=x_1-x_2$ and   used 
$${2^s-1\over 2^{s-1}}\zeta(s)=\sum_{n=1}^{\infty}{1\over n^s}+(-1)^{n-1}{1\over n^s}={1\over 2}\int_0^{\infty}{\rmd x}x^{s-1}({1\over e^x-1}+{1\over e^x+1})={1\over (s-1)!}\int_0^1{{\rmd x}\over x}(\log{1+x\over 1-x})^{s-1}$$
or rather, from (\ref{cb}),
\begin{align}\int_{0}^{\infty}\rmd u\,  u
K_{0}(u)^4
&={1\over 4} \int_{0}^{1}{\rmd x}{1\over x}(\log{1+x\over 1-x})^2={1\over 2}\int_{0}^{1}{\rmd u}{1\over 1-u^2}(\log u)^2\nonumber\\
&={1\over 2}\int_{0}^{1}{\rmd u}(\log u)^2\sum_{n=0}^{\infty}u^{2n}={7\zeta(3)\over 8}\end{align}
with the change of variable $u={1-x\over 1+x}$.
 
One knows for example that $\int_{0}^{\infty}\rmd u\,  u^3K_{0}(u)^4$ is a linear combination with rational coefficients of $1$ and $\int_{0}^{\infty}\rmd u\,  u K_{0}(u)^4$ namely
\be 4\int_{0}^{\infty}\rmd u\,  u
K_{0}(u)^4-16\int_{0}^{\infty}\rmd u\,  u^3
K_{0}(u)^4=3\ee
One has
\begin{align}\int_{0}^{\infty}\rmd u\,  u^3
K_{0}(u)^4=&\int_{0}^{1}{\rmd x_1} \int_{0}^{1-x_1}{\rmd x_2}\left({1+(x_1+x_2)^2\over 2(x_1+x_2)}\log{1+x_1+x_2\over 1-(x_1+x_2)}-1\right) {1-(x_1+x_2)^2\over (x_1+x_2)^2 }\nonumber\\
&{4x_1x_{2}\over \bigg( 1-(x_1+x_2)^2+4x_1x_{2}\bigg)^2}\end{align}
so  that
\begin{align}\label{relation} & \int_{0}^{1}{\rmd x_1} \int_{0}^{1-x_1}{\rmd x_2} \log{1+x_1+x_2\over 1-x_1-x_2}{1\over x_1+x_2} 
{2\over1-(x_1+x_2)^2+ 4 x_1x_2}\nonumber\\ 
&-8
\left({1+(x_1+x_2)^2\over 2(x_1+x_2)}\log{1+x_1+x_2\over 1-(x_1+x_2)}-1\right){1-(x_1+x_2)^2\over (x_1+x_2)^2 } 
{4x_1x_{2}\over \bigg(1-(x_1+x_2)^2+4 x_1x_{2}\bigg)^2}=3\end{align}
has to be  satisfied.
With the change of variables $x=x_1+x_2$ and $y=x_1-x_2$  
\begin{align} &\int_{0}^{1}{\rmd x_1} \int_{0}^{1-x_1}{\rmd x_2} \left({1+(x_1+x_2)^2\over 2(x_1+x_2)}\log{1+x_1+x_2\over 1-(x_1+x_2)}-1\right){1-(x_1+x_2)^2\over (x_1+x_2)^2 }{4x_1x_{2}\over \bigg(1-(x_1+x_2)^2+4 x_1x_{2}\bigg)^2} \nonumber\\
&={2\over 2 } \int_{0}^{1}{\rmd x}\left({1+x^2\over 2x}\log{1+x\over 1-x}-1\right){1-x^2\over x^2}\int_{0}^{x}{\rmd y}{(x+y)(x-y)\over(1-y^2)^2} 
\nonumber\\
&=\int_{0}^{1}{\rmd x}\left({1+x^2\over 2x}\log{1+x\over 1-x}-1\right){1-x^2\over x^2}{1\over 2}\left(-x+(1+x^2){1\over 2}\log{1+x\over 1-x}\right)\nonumber\\
&=\int_{0}^{1}{\rmd x}{1-x^2\over 2x}\left({1+x^2\over 2x}\log{1+x\over 1-x}-1\right)^2\end{align}
so that (\ref{relation}) becomes 
\be\int_0^1{\rmd x}\left({1\over x}(\log{ 1+x\over 1 - x})^2-4{1-x^2\over x} ({1+x^2\over 2x} \log{ 1+x\over 1 - x} - 1)^2 \right)=3\ee
which is indeed true.

Finally
\be\int_{0}^{\infty}\rmd u\,  u
I_0(u)K_{0}(u)^3={1\over 4}\int_0^1\rmd x_1\int_0^{1-x_1}\rmd x_2 {1\over x_1x_2(x_1+x_2)-(x_1+x_2)(1-x_1-x_2)}\ee
With the change of variables $x=x_1+x_2$ and $y=x_1-x_2$ one gets
\begin{align}\int_{0}^{\infty}\rmd u\,  u
I_0(u)K_{0}(u)^3=&{1\over 4}\int_0^1\rmd x\int_0^{x}\rmd y{4\over x}{1\over x-2+y}{1\over x-2-y}\nonumber\\
=&{1\over 2}\int_0^1\rmd x{1\over x(x-2)}\int_0^{x}\rmd y{1\over x-2+y}+{1\over x-2-y}\nonumber\\
=&{1\over 2}\int_0^1\rmd x{1\over x(x-2)}\log(1-x)\nonumber\\
=&-{1\over 2}\int_0^1\rmd u{1\over 1-u^2}\log u={3\zeta(2)\over  8}\end{align}
where $u=1-x$.

\subsection*{ 3) K-Bessel integrals summary}

 \[\int_0^{\infty}\rmd u\,u K_0(u)\,=\,1\]
 
 \[\int_0^{\infty}\rmd u\, K_0(u)\,=\,{\pi\over 2}=-{\sqrt{3}\over 2}(\psi_0(1/3)-\psi_0(2/3))\]
 
 \[\int_0^{\infty}\rmd u\,u K_0(u)^2\,=\,{1\over 2}\]
 
 \[\int_0^{\infty}\rmd u\, K_0(u)^2\,=\,{3\over 2}\zeta(2)\]
 
 \[\int_0^{\infty}\rmd u\, u K_0(u)^3\,=\,{\psi_1(1/3)-\psi_1(2/3)\over 12}\]
 
 \[\int_0^{\infty}\rmd u\,  K_0(u)^3\,=\, ? \]
 
 \[\int_0^{\infty}\rmd u\, u K_0(u)^4\,=\,{7\over 8}\zeta(3)\]
 
 \[\int_0^{\infty}\rmd u\,  K_0(u)^4\,=\,?\]
 
 \[\ldots\,?\]
 
 \[\int_0^{\infty}\rmd u\, u K_0(u)^8-{72}\int_0^{\infty}\rmd u\, u^3 K_0(u)^8=_{\rm PSLQ}77\zeta(5)\]
 
 \[\ldots\,?\]

  \[\lim_{n\to\infty}{2^{n-1}\over n!}\int_0^{\infty}\rmd u\, u K_0(u)^n=e^{2\psi_0(1)}\]
 
 \[\lim_{n\to\infty}{1\over n!}\int_0^{\infty}\rmd u\,  K_0(u)^n=2 e^{\psi_0(1)}\]
 
where $\psi_0(1)$ is minus the Euler constant.

} 
 
 \section*{ APPENDIX B: HIGHER ORDER RECURSIONS}
 
  From the iteration (\ref{rec2}) it is  easy to get for $\kappa\ge 5$ higher order recursions generalizing the  recursions (\ref{fractionbis}, \ref{fractionbister}) for $\kappa=4,3$. For example in the  $\kappa=4$ case, (\ref{rec2}) or (\ref{simple0}) becomes, defining $x(k)=  I_{2k,0}^{(4)}, \;y(k)=I_{2k,4}^{(4)}$, 
\be\label{simplebis}\left(
\begin{array}{c}
x(k)\\ 
y(k)\end{array}\right)
={1 + 2k\over (k-1)k(k+1)^2}
\left(
\begin{array}{cc}
{ b(k)} & {d(k)}\\  
{   a(k)}&{c(k)}  
\end{array}\right)
\left(
\begin{array}{c}
x(k+1)\\ 
y(k+1)\end{array}\right)
%\nonumber\ee
\ee
where $a(k), b(k), c(k), d(k)$ are given in (\ref{abcd}).

One can for example start from the inverse iteration    
\be\label{simpleter}\left(
\begin{array}{c}
x(k+1)\\ 
y(k+1)\end{array}\right)
=
\left(
\begin{array}{cc}
{\alpha(k)} & {\beta(k)}\\  
{   \gamma(k)}&{\delta(k)}  
\end{array}\right)
\left(
\begin{array}{c}
x(k)\\ 
y(k)\end{array}\right)
%\nonumber\ee
\ee
where 
\begin{align}\alpha(k)&={ (k-1)k(k+1)^2\over 1 + 2k}{1\over b(k)c(k)-a(k)d(k)}c(k)\nonumber\\
\beta(k)&=-{ (k-1)k(k+1)^2\over 1 + 2k}{1\over b(k)c(k)-a(k)d(k)}d(k)\nonumber\\
\gamma(k)&=-{ (k-1)k(k+1)^2\over 1 + 2k}{1\over b(k)c(k)-a(k)d(k)}a(k)\nonumber\\
\delta(k)&={ (k-1)k(k+1)^2\over 1 + 2k}{1\over b(k)c(k)-a(k)d(k)}b(k)\label{needed}\end{align} 

Consider $k\to k+1$ in (\ref{simpleter}) to get 
\be x(k+2)={\alpha(k+1)x(k+1)+ \beta(k+1)y(k+1)}\label{almost}\ee
Invert (\ref{simpleter})    
\be x(k)={1\over \alpha(k)\delta(k)-\beta(k)\gamma(k)} (\delta(k)x(k+1)- \beta(k)y(k+1))\label{notend}\ee
and use (\ref{notend}) to eliminate $y$ in (\ref{almost}). One  obtains a recursion for $x$
\be\beta(k)x(k+2)-(\beta(k)\alpha(k+1)+\beta(k+1)\delta(k))x(k+1)+(\alpha(k)\delta(k)-\beta(k)\gamma(k))\beta(k+1)x(k)=0\ee
that is to say, using (\ref{needed})
\be k^4 x(k-1) -  (2k-1) (2k+1) (2 + 5k + 5k^2) x(k) + 
 16 (1 + k)^2 x(k+1)=0\label{lastofthelast}\ee
When multiplied by $16k^2$ the recursion (\ref{lastofthelast})  is (\ref{fractionbis}) for ${\tilde{x}}(k)= 8^k (2k)! k!x(k)$.

In the $\kappa=5$  case, the iteration  (\ref{rec2}) for  $x(k)=I_{2k,0}^{(5)}$, $y(k)=I_{2k,2}^{(5)}$ and $z(k)=I_{2k,4}^{(5)}$ 
is  
\be\label{simple5}\left(
\begin{array}{c}
x(k)\\ 
y(k)\\z(k)\end{array}\right)
=
\left(
\begin{array}{ccc}
 {5(1 + 2k)\over 2(k+1)} & {10(1 + 2k)\over (k+1)}&0\\  
 {1+2k\over k}&{(1+2k)(9+17k)\over 2k^2}&{3(1+k)(1+2k)\over k^2}\\ 
{0}&{6(1+k)(1+2k)\over k(k-1)}&{(1+k)(1+2k)(-8+13k)\over2k(k-1)^2} 
\end{array}\right)
\left(
\begin{array}{c}
x(k+1)\\ 
y(k+1)\\z(k+1)\end{array}\right)
%\nonumber\ee
\ee

Again one may start from the inverse iteration 

\be\label{simplequar}\left(
\begin{array}{c}
x(k+1)\\ 
y(k+1)\\z(k+1)\end{array}\right)
=
\left(
\begin{array}{ccc}
{\alpha(k)} & {\beta(k)}&\gamma(k)\\  
{\delta(k)}&{\epsilon(k)} &{\zeta(k)}\\
{\eta(k)}&{\theta(k)}&{\iota(k)}  
\end{array}\right)
\left(
\begin{array}{c}
x(k)\\ 
y(k)\\z(k)\end{array}\right)
%\nonumber\ee
\ee
where in particular $\beta(k)\zeta(k)-\gamma(k)\epsilon(k)=0$ and $\delta(k)\theta(k)-\epsilon(k)\eta(k)=0$ to account  for the vanishing elements of (\ref{simple5}).  In (\ref{simplequar})  consider   $k\to k+1$ 
\be \label{11} z(k+2)={\eta(k+1)}x(k+1)+{\theta(k+1)}y(k+1)+{\iota(k+1)}z(k+1)  \ee 
\be \label{22} y(k+2)={\delta(k+1)}x(k+1)+{\epsilon(k+1)}y(k+1) +{\zeta(k+1)z(k+1)}\ee 
and $k\to k+2$ 
\be \label{33}y(k+3)={\delta(k+2)}x(k+2)+{\epsilon(k+2)}y(k+2) +{\zeta(k+2)z(k+2)}\ee

(\ref{22}, \ref{33}) allow to eliminate the $z $'s in (\ref{11}) 
\begin{align}  &{1\over \zeta(k+2)} ( y(k+3)-{\delta(k+2)}x(k+2)-{\epsilon(k+2)}y(k+2) )\nonumber\\&={\eta(k+1)}x(k+1)+{\theta(k+1)}y(k+1)+{\iota(k+1)\over \zeta(k+1)} ( y(k+2)-{\delta(k+1)}x(k+1)-{\epsilon(k+1)}y(k+1) ) \label{44}\end{align}
 Invert (\ref{simplequar}) 
\be \label{55}
x(k)={\left(-\zeta(k)\theta(k) + \epsilon(k)\iota(k)\right)x(k+1)+\left(\gamma(k) \theta(k) - \beta(k) \iota(k)\right)y(k+1)\over \gamma(k)\delta(k) \theta(k)- \alpha(k) \zeta(k)  \theta(k)- \beta(k)\delta(k) \iota(k) + \alpha(k) \epsilon(k) \iota(k)}
\ee 
and use (\ref{55}) to eliminate the $y$'s in (\ref{44}). One finally obtains  a recursion for $x$, which reads, using the matrix elements in (\ref{simple5}) 
\begin{align}  &8 k^5 x(k-1) - (-1 + 
    2 k) \bigg(4 \left(5 +28k + 63k^2 + 70k^3 + 35k^4\right) x(k) \nonumber\\&- (1+k) (1+2 k) \left(2 \left(285+518k+259k^2\right)x(k+1) - 
       225 (2 + k) (3 + 2k) x(k+2)\right)\bigg)=0\label{66}\end{align} 
Generalizing this construction to higher order recursions for  $\kappa>5$ is straightforward -see \cite{salve} for recursions of this type 
  (see also \cite{zu2} for an example involving $\zeta(5)$ and $\zeta(3)$).

\section*{ APPENDIX C: A FAMILY OF DOUBLE NESTED INTEGRALS}

Coming back to (\ref{wellbis},\ref{PSLQ}) use \[\int_0^{\infty}g(u)\rmd u \int_u^{\infty} f(x)\rmd x =\int_0^{\infty}f(u)\rmd u \int_0^u g(x)\rmd x\] to rewrite $I_{\rho^2\alpha^6}$ as
\begin{align} \label{wellter}
I_{\rho^2\alpha^6} =  8 & \int_0^{\infty}\rmd u \,u \,K_0(u)^2(uK_1(u))^2
\int_0^{u}\rmd x\,xK_1(x)I_1(x)K_0(x)^2 \nonumber \\
-4&\int_0^{\infty}\rmd u\, u\,K_0(u)^2K_1(u)^2 \int_0^{u}\rmd x\, x K_0(x)xK_1(x)\big(xK_1(x)I_0(x)-xK_0(x)I_1(x)\big)
\nonumber\\ 
+&\int_0^{\infty}u\,K_0(u)^4(uK_1(u))^2 \rmd u 
\end{align}
Next use $xI_1(x)K_0(x)+xI_0(x)K_1(x)=1 $ so that 
\begin{align} \label{wellterter}
I_{\rho^2\alpha^6} =  8 &\int_0^{\infty}\rmd u \,u^3 \,K_0(u)^2K_1(u)^2
\int_0^{u}\rmd x\,xK_1(x)I_1(x)K_0(x)^2 \nonumber \\
 +8 &\int_0^{\infty}\rmd u \,u \,K_0(u)^2K_1(u)^2
\int_0^{u}\rmd x\,x^3K_1(x)I_1(x)K_0(x)^2 \nonumber \\
 -4&\int_0^{\infty}\rmd u\, u\,K_0(u)^2K_1(u)^2 \int_0^{u}\rmd x\, x^2 K_0(x)K_1(x)
\nonumber\\ 
+&\int_0^{\infty}u\,K_0(u)^4(uK_1(u))^2 \rmd u 
\end{align}
One has for the next to last  term in (\ref{wellterter})
  \begin{align} \label{wellquar}
 -4\int_0^{\infty}\rmd u\, u\,K_0(u)^2K_1(u)^2 \int_0^{u}\rmd x\, x^2 K_0(x)K_1(x) =2 &\int_0^{\infty}\, u\,K_0(u)^2K_1(u)^2 \big((u K_1(u))^2-1\big)\rmd u \nonumber\\
 =-{2\over 3}&\int_0^{\infty} u\, K_1(u) \,\big((u K_1(u))^2-1\big)\rmd K_0(u)^3\nonumber \\
={2\over 3}&\int_0^{\infty}   K_0(u)^3\,\rmd (u K_1(u))\big((u K_1(u))^2-1\big) \nonumber\\
={2\over 3}&\int_0^{\infty} u K_0(u)^4  \rmd u\, -2\int_0^{\infty}u\,K_0(u)^4(uK_1(u))^2 \rmd u
\end{align}
so that
\begin{align} \label{wellquarquar}
I_{\rho^2\alpha^6} = 8 &\int_0^{\infty}\rmd u \,u^3 \,K_0(u)^2K_1(u)^2
\int_0^{u}\rmd x\,xK_1(x)I_1(x)K_0(x)^2 \nonumber \\
+8 &\int_0^{\infty}\rmd u \,u \,K_0(u)^2K_1(u)^2
\int_0^{u}\rmd x\,x^3K_1(x)I_1(x)K_0(x)^2 \nonumber \\
+{2\over 3}&\int_0^{\infty} u K_0(u)^4  \rmd u
\nonumber\\ 
-&\int_0^{\infty}u\,K_0(u)^4(uK_1(u))^2 \rmd u 
\end{align}

Defining  
\[f_n(x) = x^n K_0(x)^2 K_1(x)^2\quad {\rm and} \quad g_n(x) = x^n K_0(x)^2 K_1(x) I_1(x)\]
\noindent equations (\ref{first}, \ref{rel},   \ref{wellquarquar}) imply that 
%$ \zeta(f_3,g_1)+\zeta(f_1,g_3)$ 
%&=&  \int_0^{\infty}\rmd u \,u^3 \,K_0(u)^2K_1(u)^2
%\int_0^{u}\rmd x\,xK_1(x)I_1(x)K_0(x)^2    \nonumber \\&+& \int_0^{\infty}\rmd u \,u \,K_0(u)^2K_1(u)^2
%\int_0^{u}\rmd x\,x^3K_1(x)I_1(x)K_0(x)^2\eea
\begin{align} \label{nice}{\tilde \zeta}(f_3,g_1)+{\tilde \zeta}(f_1,g_3)=_{PSLQ}&{1\over 48}\int_{0}^{\infty}u \, K_0(u)^6\,\rmd u
- {3\over 160}\int_{0}^{\infty}u^3 K_0(u)^6\,\rmd u\nonumber\\&- {7\over 96}\,\zeta(3)-{31\over 1280}\,\zeta(5)
\end{align}
as confirmed by a direct  PSLQ check.  

So the   meaning of (\ref{PSLQ}) when rewritten as (\ref{nice}) might be that it is the symmetric form  ${\tilde \zeta}(f_3,g_1)+{\tilde \zeta}(f_1,g_3)$ which is a linear combination with rational coefficients of simple Bessel integrals of weigth $6$ and $\zeta$ numbers like $\zeta(5)$ and below. This suggests to look at other  symmetric sums of double nested integrals sharing this property. A PSLQ search confirms that
${\tilde \zeta}(f_5,g_1)+{\tilde \zeta}(f_1,g_5)$ belongs indeed to this category 
\begin{align} {\tilde \zeta}(f_5,g_1)+{\tilde \zeta}(f_1,g_5)&=_{PSLQ}{211\over 11520}\int_{0}^{\infty}u \, K_0(u)^6\,\rmd u
+ {3953\over 23040}\int_{0}^{\infty}u^3 K_0(u)^6\,\rmd u\nonumber\\&+{11\over 9216}- {1\over 9}\,\zeta(3)-{93\over 5120}\,\zeta(5)
\end{align}
%$-(211/11520), -(3953/23040), -(11/9216), 1/9, 93/5120, 1$
as well as ${\tilde \zeta}(f_7,g_1)+{\tilde \zeta}(f_1,g_7)$
  \begin{align} {\tilde \zeta}(f_7,g_1)+{\tilde \zeta}(f_1,g_7)&=_{PSLQ}{108731\over 1728000}\int_{0}^{\infty}u \, K_0(u)^6\,\rmd u
+{4256617\over 3456000}\int_{0}^{\infty}u^3 K_0(u)^6\,\rmd u\nonumber\\&+{27877\over 
  460800}- {8\over 15}\,\zeta(3)-{279
  \over 5120}\,\zeta(5)
\end{align}
%$-(108731/1728000), -(4256617/3456000), -(27877/
%  460800), 8/15, 279/5120, 1$
  and ${\tilde \zeta}(f_3,g_5)+{\tilde \zeta}(f_5,g_3)$
\begin{align} {\tilde \zeta}(f_3,g_5)+{\tilde \zeta}(f_5,g_3)&=_{PSLQ}-{28921\over 691200}\int_{0}^{\infty}u \, K_0(u)^6\,\rmd u
+ {1151533\over 1382400}\int_{0}^{\infty}u^3 K_0(u)^6\,\rmd u\nonumber\\&+{14653\over 184320}+ {25\over 192}\,\zeta(3)+{279\over 20480}\,\zeta(5)
\end{align}
%$-(28921/691200), 1151533/1382400, 14653/184320, 25/192, 279/20480, -1$ 

  We infer that  for any two odd positive integers $n,m$ one should have that ${\tilde \zeta}(f_n,g_m)+{\tilde \zeta}(f_m,g_n)$ can be rewritten as a linear combination of  $1,\; \int_{0}^{\infty}u \, K_0(u)^6\,\rmd u,\;\int_{0}^{\infty}u^3 \, K_0(u)^6\,\rmd u$ and $\zeta(3)$ and $\zeta(5)$. Finding the coefficients of the linear combination can be in principle obtained by generalizing integration by part procedures  in \cite{nous,  nousbis, moi, mash} to double nested integrals. 
  
  \pagebreak

\end{document}